\documentclass[aps,twocolumn,superscriptaddress,longbibliography]{revtex4-1} 
\usepackage[english]{babel}
\usepackage{amsmath, bm, amssymb}
\usepackage{mathtools}
\usepackage{siunitx}
\usepackage{environ}
\usepackage{mathrsfs}
\usepackage{braket}
\usepackage[]{hyperref}
\DeclareMathAlphabet{\mathscrlower}{OT1}{pzc}{m}{it} 
\usepackage{prettyref}
\usepackage{graphicx}
\usepackage{booktabs}
\usepackage{array}
\usepackage{multirow}
\usepackage{dcolumn}
\usepackage{threeparttable}
\usepackage{threeparttablex}
\usepackage{placeins}
\usepackage[stable]{footmisc}
\usepackage{mhchem}
\usepackage{xfrac}
\usepackage{xcolor}

\newcommand{\pauli}{\boldsymbol{\sigma}}
\newcommand{\Pauli}{\boldsymbol{\sigma}}

\newcommand{\diraca}{\vec{\boldsymbol{\alpha}}}

\newcommand{\diracb}{\boldsymbol{\beta}}

\newcommand{\momop}{\hat{\vec{p}}}

\newcommand{\spinmom}{\vec{\Pauli}\cdot\momop}

\newcommand{\Sum}[2]{\sum\limits_{#1}^{#2}}

\newcommand{\parantheses}[1]{\left(#1\right)}
\newcommand{\brackets}[1]{\left[#1\right]}

\let\nablatmp\nabla
\renewcommand{\nabla}{\vec{\nablatmp}}
\DeclarePairedDelimiter\abs{\lvert}{\rvert}
\makeatletter
\let\oldabs\abs
\def\abs{\@ifstar{\oldabs}{\oldabs*}}

\newcommand{\partiell}[2]{\frac{\partial #1}{\partial #2}}
\AtBeginDocument{
\heavyrulewidth=.08em
\lightrulewidth=.05em
\cmidrulewidth=.03em
\belowrulesep=.65ex
\belowbottomsep=0pt
\aboverulesep=.4ex
\abovetopsep=0pt
\cmidrulesep=\doublerulesep
\cmidrulekern=.5em
\defaultaddspace=.5em
}

\graphicspath{ {./pictures/} }

\AtBeginDocument{%
  \newwrite\bibnotes
  \def\bibnotesext{Notes.bib}
  \immediate\openout\bibnotes=\jobname\bibnotesext
  \immediate\write\bibnotes{@CONTROL{REVTEX41Control}}%
  \immediate\write\bibnotes{@CONTROL{apsrev41Control,author="08",editor="1",pages="1",title="0",year="1"}}%
}

\begin{document}
\title{Enhanced sensitivity to variations of fundamental constants in
highly charged molecules from analytic perturbation theory}
\date{\today}
\author{Carsten Z\"ulch}
\affiliation{Fachbereich Chemie, Philipps-Universit\"{a}t Marburg, Hans-Meerwein-Stra\ss{}e 4, 35032 Marburg}
\author{Konstantin Gaul}
\affiliation{Fachbereich Chemie, Philipps-Universit\"{a}t Marburg, Hans-Meerwein-Stra\ss{}e 4, 35032 Marburg}
\author{Robert Berger}
\email[]{robert.berger@uni-marburg.de}
\affiliation{Fachbereich Chemie, Philipps-Universit\"{a}t Marburg, Hans-Meerwein-Stra\ss{}e 4, 35032 Marburg}
\begin{abstract}
Quasi-forbidden electronic transitions in atoms and vibronic
transitions between quasi-degenerate states in molecules serve as
powerful probes of hypothetical temporal variations of fundamental
constants. Computation of the sensitivity of a transition to a
variation of the fine-structure constant is conventionally performed by
numerical variation of the speed of light in sophisticated electronic
structure calculations, and therewith several individual calculations
have to be performed. An approach is presented herein that obtains
sensitivity coefficients as perturbation to the
Dirac-Coulomb Hamiltonian and allows, for instance, computation of their
leading-order contributions as expectation
values of the relativistic kinetic energy and rest-mass operators.
These are available in essentially all \emph{ab initio} relativistic electronic structure codes.
Additionally, the corresponding operators for two-component
Hamiltonians are derived, explicitly for the zeroth order regular approximation
Hamiltonian. The approach is applied to demonstrate great
sensitivity of highly charged polar molecules that were recently
proposed for high-precision spectroscopy in [Z\"ulch \emph{et al.},
arXiv:2203.10333[physics.chem-ph]]. In particular, a
high sensitivity of a wealth of quasi-degenerate vibronic transitions in \ce{PaF^3+}
and \ce{CeF^2+} to temporal variations of the fine-structure constant
and the electron-proton mass ratio is shown.
\end{abstract}

\maketitle

\section{Introduction}%
Theories unifying gravity and quantum mechanics predict temporal variations of
fundamental constants (VFC), such as the fine-structure constant $\alpha$ and the
electron-proton mass-ratio $\mu=m_\mathrm{e}/m_\mathrm{p}$, with the expansion
of our universe (see Review~\cite{uzan:2003}; we consider fundamental
physical constants inherently free of units; for a discussion see e.g. Refs.~\cite{duff:2002, duff:2015}). 
Moreover, such variations could arise from dependence of
$\alpha$ and $\mu$ on the gravitational potential and can be a signature of
scalar bosonic dark matter like dilatons. Precision
spectroscopy of atoms and molecules gave one of the most stringent limits of the
variations of $\alpha$ and $\mu$ \cite{uzan:2003}. The so far best upper bound on a
temporal variation of these two dimensionless quantities was obtained by comparing two
transition frequencies in \ce{Yb+} \cite{godun:2014,huntemann:2014,lange:2021}:
$\delta\mu/\mu<10^{-17}\,\si{\per yr}$ and
$\delta\alpha/\alpha<10^{-18}\,\si{\per yr}$, where a 10-fold improvement is
expected in near future \cite{safronova:2018}. Heavy-element-containing
diatomic molecular ions
\cite{flambaum:2007,demille:2008a,beloy:2010,beloy:2011,beloy:2011a,pasteka:2015}
and highly charged atoms are promising candidates for improvement of these
bounds as per their much higher sensitivity  due to strong relativistic effects and
transitions that have very small linewidths \cite{berengut:2010,berengut:2011}.

For identifying transitions that appear favourable for detecting VFC, atomic and
molecular theory is required. To the best of our knowledge, in all previous theoretical studies of VFC,
the electronic enhancement of an $\alpha$ variation was determined by numerically varying the
speed of light in relativistic electronic structure calculations. This requires
to perform an array of possibly quite demanding computations to
obtain the sensitivity to temporal variations of $\alpha$. 

This tedious 
numerical procedure for obtaining enhancement factors, which becomes particular 
costly for larger molecular systems or for high-level many-body approaches, can likely be considered as one of the
reasons for the relatively low number of (\textit{ab initio}) predictions presently available.
Predictions for atomic systems are somewhat more abundant (see e.g. \cite{dzuba:1999,safronova:2018}),
whereas for molecular systems information is scarce
\cite{flambaum:2006,flambaum:2013,flambaum:2007,pasteka:2015,beloy:2011b} (see also
Ref.~\cite{safronova:2019}, \cite{flambaum:2008}, and \cite{zelevinsky:2008} as well as references
therein). 

In this paper we
show that within relativistic four-component approaches, the sensitivity to
a small variation of the fine-structure constant can be calculated directly as
expectation value of the Dirac kinetic energy and electronic rest-mass
operators scaled by a constant factor. We demonstrate the applicability for
four-component and quasi-relativistic two-component calculations, discuss
higher-order effects and demonstrate the consistency by comparison to the Pauli
expansion. The method proposed is directly amenable to common relativistic
electronic structure programs such as Dirac \cite{dirac19} as the perturbing
operators needed are already available.

We utilise this analytic approach to compute the VFC sensitivity of stable polar
highly charged molecular ions such as \ce{PaF^3+}, which we recently suggested
as prospective laboratories for precision tests of fundamental physics
\cite{zulch:2022}. Such systems possess close-lying electronic states of different
symmetry and exhibit large relativistic effects. Thus, we anticipated them to
be ideal candidates to search for VFC \cite{zulch:2022,zulch:2023}. Furthermore,
we study \ce{CeF^2+} alongside \ce{PaF^3+}, as both ions feature similar electronic
characteristics for the lower electronic states. We demonstrate a high sensitivity of both highly
charged species to variations of $\alpha$ and $\mu$ by application of our
novel approach. 

\section{Theory}
\subsection{Influence of variations of $\alpha$ and $\mu$ on vibronic
transition frequencies}
Assuming that the fine-structure constant $\alpha$ varies by a small amount
$\delta\alpha$ we can define the red-shifted fine-structure
constant $\alpha_z$ as $\alpha_z=\alpha+\delta\alpha$.
The electronic transition wavenumber in an atom or molecule
changes
with the red-shifted fine structure constant as \cite{webb:1999,dzuba:1999}
\begin{multline}
\tilde{\nu}= \tilde{\nu}_{\mathrm{el}} +
\Delta q_{1} \parantheses{\parantheses{\frac{\alpha_z}{\alpha}}^2-1} \\+
\Delta q_{2}' \parantheses{\parantheses{\frac{\alpha_z}{\alpha}}^4-1}
+\mathcal{O}(\alpha_z^6)\,.
\label{eq: shift_trad}
\end{multline} 
Here $\Delta q_{1}$ and $\Delta q_{2}'$ describe the sensitivity to a variation of the
fine-structure constant and need to be determined from electronic
structure theory. 
The expansion coefficients $q_1$ and $q_2'$ indicate for a given electronic state
its sensitivity to a change of the fine structure constant and
have previously been
determined numerically by changing $\alpha$ in the electronic structure calculations for each electronic state. $\Delta
q_i$ is then computed from the difference $q_i^{\mathrm{state}\ m} - q_i^{\mathrm{state}\ n}$, where $m$ and $n$ are the
electronic states involved in the transition $m \leftarrow n$ with corresponding transition wavenumber
$\tilde{\nu}_\mathrm{el}$.
As $\delta\alpha$ is expected to be as small as 
$\delta\alpha/\alpha\lesssim10^{-18}$ \cite{lange:2021}, $q_2'$ currently has no practical relevance
for experimental physics and therefore serves for the time being primarily as a test of consistency of the method
derived herein.
We will use a slight modification of the definition of $q_{2}'$ and write
(now with $q_{2}$ instead of $q_{2}'$)
\begin{equation}
\tilde{\nu}=  \tilde{\nu}_{\mathrm{el}} +
\Delta q_{1} \tilde{x} + \Delta q_{2} \tilde{x}^2 + \mathcal{O}(\tilde{x}^3)\,.
\label{eq: shift_here}
\end{equation}
where we used the variable 
\begin{equation}
\tilde{x}=\parantheses{\frac{\alpha+\delta\alpha}{\alpha}}^2-1=\frac{2\delta\alpha}{\alpha}+\frac{\delta\alpha^2}{\alpha^2}\,.
\label{eq: alphadep}
\end{equation}
This formulation, which was also chosen in
Ref.~\cite{hansmann:2020}, is consistent with the overall
$\alpha$-dependence in relativistic electronic structure theory and
coincides with the Pauli expansion as we demonstrate later.

A change, $\delta\tilde{\nu}_{v}$, in the vibronic transition wavenumber $\tilde{\nu}_{v}$ of a molecule,
where the subscript $v$ indicates the vibrational transition, due to a
temporal variation $\delta\alpha$ of $\alpha$, and $\delta\mu$ of $\mu$
reads in leading order\cite{flambaum:2007} as
\begin{equation}
\delta\tilde{\nu}_{v} = \Delta q_{1}\tilde{x} +
\frac{\tilde{\nu}_{\mathrm{vib}}}{2}
\frac{\delta\mu}{\mu}\,,
\label{eq: shift_leading_order}
\end{equation}
where $\tilde{x}\sim2\delta\alpha/\alpha$ and
$\tilde{\nu}_{\mathrm{vib}}$
is the pure vibrational transition wavenumber for transition $v$. 
Note that we take anharmonically computed term values, but account only for the leading order dependence
of the energy levels on the reduced mass of the nuclei. The latter gives rise to the factor of 1/2.

The overall enhancements $K_{\alpha,v}$, $K_{\mu,v}$ of variations of fundamental constants
$\alpha$ and $\mu$ is found from the leading-order dependence of the transition
wavenumber $\tilde{\nu}_{v}$ on the fundamental constants
\cite{flambaum:2007}
\begin{equation}
\frac{\delta\tilde{\nu}_{v}}{\tilde{\nu}_{v}} \approx
\frac{2 \Delta q_{1}}{\tilde{\nu}_{v}}\frac{\delta\alpha}{\alpha}
+\frac{\tilde{\nu}_{\mathrm{vib}}}{2\tilde{\nu}_{v}}\frac{\delta\mu}{\mu}
=\left(K_{\alpha,v}\frac{\delta\alpha}{\alpha}
+K_{\mu,v}\frac{\delta\mu}{\mu}\right)\,.
\end{equation}

\subsection{Variation of $\alpha$ as small perturbation to the
Dirac--Coulomb Hamiltonian}
We start from the Dirac--Coulomb (DC) Hamiltonian in which the potential energy
operator is $\alpha$ independent. Additional relativistic corrections to the
potential energy of order $\alpha^2$, such as those caused by the Breit interaction, can become
relevant in particular for light elements but are expected to be negligible for
heavy-elemental systems such as \ce{CeF^2+} and \ce{PaF^3+}. When quadratic
effects ($q_2$) are desired, leading order quantum electrodynamics (QED) corrections
could contribute to these as well (see Ref.~\cite{janke:2025} for our recent approaches
to incorporate QED corrections within a two-component framework), but those are neglected in our present work. 

The electronic one-particle DC-Hamiltonian reads
\begin{equation}
\hat{H}_\mathrm{DC} = c \diraca\cdot\momop +
\parantheses{\diracb-\bm{1}} m_\mathrm{e} c^2 + \hat{V}_\mathrm{C}\,,
\end{equation} 
where we shifted the energy spectrum by $m_\mathrm{e} c^2$, with $c$ being
the speed of light, and
used the one-electron linear momentum operator
$\hat{\vec{p}}=-i\hbar\nabla$, the instantaneous, $\alpha$-independent Coulomb interaction
$\hat{V}_\mathrm{C}$, the standard Dirac matrices $\diraca$,
$\diracb$ and the $4\times4$
unit matrix $\bm{1}$.
In atomic units, $m_\mathrm{e}$ has the numerical value of 1 and $c$ the
numerical value of $\alpha^{-1}$ yielding
\begin{equation}
\hat{H}_\mathrm{DC} = \alpha^{-1} \diraca\cdot\momop +
\parantheses{\diracb-\bm{1}} \alpha^{-2} + \hat{V}_\mathrm{C}\,.
\end{equation}
As mentioned above, possible variations $\delta\alpha$ in the fine structure constant
are expected to be on the order
$\delta\alpha/\alpha\lesssim10^{-18}$ allowing the use of first order
perturbation theory. For such a small variation in $\alpha$ we set
$\alpha\rightarrow\alpha+\delta\alpha$ yielding the perturbed
DC-Hamiltonian 
\begin{equation}
\hat{\tilde{H}}_\mathrm{DC} = \frac{1}{\alpha+\delta\alpha} \diraca\cdot\momop +
\parantheses{\diracb-\bm{1}}
\frac{1}{\parantheses{\alpha+\delta\alpha}^{2}} + \hat{V}_\mathrm{C}\,.
\end{equation}
To find the leading order perturbations in $\delta\alpha$ for the
DC-Hamiltonian, we use following expansion employing the definition
(\ref{eq: alphadep}):
\begin{align}
\frac{1}{\alpha+\delta\alpha}&=\alpha^{-1}\frac{1}{1+\delta\alpha/\alpha}=\alpha^{-1}\frac{1}{\sqrt{1+\tilde{x}}}\nonumber\\
&=\alpha^{-1}\Sum{k=0}{\infty}\brackets{\frac{(2k)!}{(k!)^2}\parantheses{-\frac{\tilde{x}}{4}}^k}\\
\frac{1}{\parantheses{\alpha+\delta\alpha}^2}&=\alpha^{-2}\frac{1}{1+\tilde{x}}
=\alpha^{-2}\Sum{k=0}{\infty}\brackets{-\tilde{x}}^k
\end{align}
Under a variation in the fine-structure constant the DC-Hamiltonian, thus,
takes the form
\begin{equation}
\begin{aligned}
\hat{\tilde{H}}_\mathrm{DC} 
&= \underbrace{c\diraca\cdot\momop+
\hat{V}_\mathrm{C} + \parantheses{\diracb-\bm{1}}
c^2}_{\hat{H}_\mathrm{DC}} \\
&+\underbrace{\Sum{k=1}{\infty}\brackets{\parantheses{-\tilde{x}}^k\parantheses{
\frac{(2k)!}{(k!)^24^k}c\diraca\cdot\momop+
\parantheses{\diracb-\bm{1}} c^2}
}}_{\hat{H}_\mathrm{DC}(\delta\alpha)}
\end{aligned}
\end{equation}
As $\delta\alpha$ is very small, first order perturbation theory is
sufficient for practical purposes and we obtain
the perturbing operator
\begin{equation}
\hat{H}^{(1)}_\mathrm{DC}
=\left.\partiell{\hat{H}_\mathrm{DC}(\delta\alpha)}{\tilde{x}}\right|_{\tilde{x}=0}
= -\frac{c}{2}\diraca\cdot\momop - \parantheses{\diracb-\bm{1}} c^2
\label{eq: dirac_1order}
\end{equation}
This operator is just the sum of the rest-mass and kinetic energy operators
of the DC-Hamiltonian with a different prefactor, which are
available in essentially any relativistic quantum chemistry program package. Thus,
it is straightforward to obtain $q_1$ as the expectation value (with $hc_0$
as a conversion factor from wavenumber units to energy units)
\begin{equation}
hc_0 q_1=\Braket{\hat{H}^{(1)}_\mathrm{DC}}\,.
\label{eq: first_order_alpha_var}
\end{equation} 
If there is a need to know $q_2$, it can be obtained
within second order perturbation theory as
\begin{align} 
hc_0 q_2 = \Braket{\hat{H}^{(2)}_\mathrm{DC}} +
\Sum{i\ne0}{}\frac{\abs{\Braket{\Psi_0|\hat{H}^{(1)}_\mathrm{DC}|\Psi_i}}^2}{E_0-E_i}\,,
\label{eq: second_order_alpha_var}
\end{align} 
where we have used for notational brevity a sum-over-states formulation.
The sum runs over electronic states $i$, except $0$, of energies $E_i$
with wave functions $\Psi_i$, whereas $\Psi_0$ and $E_0$ denote the wave function and
energy of the targeted reference state. The second order Hamiltonian is 
\begin{align}
\hat{H}^{(2)}_\mathrm{DC}
= \frac{1}{2} \left.\partiell{^2\hat{H}_\mathrm{DC}(\delta\alpha)}{\tilde{x}^2}\right|_{\tilde{x}=0}
=\frac{3}{8} c\diraca\cdot\momop + \parantheses{\diracb-\bm{1}} c^2
\label{eq: dirac_2order}\,, 
\end{align} 
which also contains only rest-mass and kinetic energy operators with a different weighting. 

\subsection{Quasi-relativistic Hamiltonians}
In a quasi-relativistic theory the small component $\psi^\mathrm{S}_\mathrm{D}$ of the
DC bi-spinor is not treated
explicitly but can be computed in good approximation from the
one-particle two-component wave function as
$\psi_\mathrm{D}^\mathrm{S}\approx\psi_\mathrm{2c}^\mathrm{S}=
c\parantheses{2c^2\bm{1}+E\bm{1}-\bm{\mathcal{V}}}^{-1}\spinmom\psi_\mathrm{2c}^\mathrm{L}$,
where we assume no external fields. We introduced the general non-diagonal
two-component potential $\bm{\mathcal{V}}$ to accommodate a
common formalism before giving explicit expressions for different flavours of
quasi-relativistic
methods below.  
The approximate methods differ in the way
$c\parantheses{2c^2\bm{1}+E\bm{1}-\bm{\mathcal{V}}}^{-1}$ is expanded.
To account for these varying schemes, we expand the relativistic corrections as
\begin{equation}
\frac{\bm{1} E - \bm{\mathcal{V}}}{2c^2} = \frac{\bm{\mathcal{R}}}{2c^2}=\frac{2c^2\bm{1}+\bm{\mathcal{R}}^{(0)}}{2c^2}\tilde{\bm{\mathcal{R}}}^{-1}-\bm{1}\,.
\end{equation}
Here $\bm{\mathcal{R}}^{(0)}$ is both $\alpha$ and energy independent, 
whereas $\bm{\tilde{\mathcal{R}}}$ contains all $\alpha$ and energy
dependence. Therefore we can write 
\begin{equation}
\psi_\mathrm{2c}^\mathrm{S}=
c\parantheses{2c^2\bm{1}+\bm{\mathcal{R}}^{(0)}}^{-1}\bm{\tilde{\mathcal{R}}}\spinmom\psi_\mathrm{2c}^\mathrm{L}\, .
\end{equation}

This separation
allows for a simultaneous treatment of e.g. the Pauli expansion and a regular approximation ansatz,
where for zeroth order Hamiltonians
$\bm{\tilde{\mathcal{R}}}$ reduces to $\bm{1}$. The
quasi-relativistic Hamiltonian operators may then be written in the general form
\begin{equation}
\hat{H}_\mathrm{2c}=c^2 \spinmom
\parantheses{2c^2\bm{1}+\bm{\mathcal{R}}^{(0)}}^{-1} \bm{\tilde{\mathcal{R}}}\spinmom +
\hat{V}_{\mathrm{C}}\,.
\end{equation}
Note that methods relying on a resolution-of-the-identity expansion, like the so-called
exactly transformed two-component Hamiltonians (X2C), can principally also start from this
formulation. The direct connection may become more apparent, if a symmetric elimination
of the small component is considered leading to the Hamiltonian
\begin{align}
\begin{split}
\hat{H}_\mathrm{2c,sym}&=\hat{V}_{\mathrm{C}} + c^2 \spinmom \brackets{(\bm{\tilde{\mathcal{R}}})^\dagger \parantheses{2c^2\bm{1}+\bm{\mathcal{R}}^{(0)}}^{-1} \right. \\
      &+ \left. \parantheses{2c^2\bm{1}+\bm{\mathcal{R}}^{(0)}}^{-1} \bm{\tilde{\mathcal{R}}}}\spinmom \\
      &+ \spinmom (\bm{\tilde{\mathcal{R}}})^\dagger  \parantheses{2c^2\bm{1}+\bm{\mathcal{R}}^{(0)}}^{-1} \\
      &\times \hat{V} \parantheses{2c^2\bm{1}+\bm{\mathcal{R}}^{(0)}}^{-1} \bm{\tilde{\mathcal{R}}}\spinmom\,.
\end{split}
\end{align}
Under a variation of the fine-structure constant the 2c
Hamiltonians become
\begin{equation}
\hat{\tilde{H}}_\mathrm{2c}= \spinmom
\parantheses{2\times\bm{1}+(\alpha+\delta\alpha)^2\bm{\mathcal{R}}^{(0)}}^{-1}
\bm{\tilde{\mathcal{R}}} \spinmom +
\hat{V}_{\mathrm{C}}\,.
\label{eq: gen_2c_alphavar}
\end{equation}
and 
\begin{align}
\begin{split}
\hat{H}_\mathrm{2c,sym}&=\hat{V}_{\mathrm{C}} + \spinmom \brackets{(\bm{\tilde{\mathcal{R}}})^\dagger \parantheses{2\times\bm{1}+(\alpha+\delta\alpha)^2\bm{\mathcal{R}}^{(0)}}^{-1} \right. \\
  &+ \left.\parantheses{2\times\bm{1}+(\alpha+\delta\alpha)^2\bm{\mathcal{R}}^{(0)}}^{-1} \bm{\tilde{\mathcal{R}}}}\spinmom \\
  &+ (\alpha+\delta\alpha)^4\spinmom (\bm{\tilde{\mathcal{R}}})^\dagger  \parantheses{2\times\bm{1}+(\alpha+\delta\alpha)^2\bm{\mathcal{R}}^{(0)}}^{-1} \\
  &\times \hat{V} \parantheses{2\times\bm{1}+(\alpha+\delta\alpha)^2\bm{\mathcal{R}}^{(0)}}^{-1} \bm{\tilde{\mathcal{R}}}\spinmom\,.
\end{split}
\end{align}
Here it must be remembered that $\bm{\tilde{\mathcal{R}}}$ also 
depends on the variation of $\alpha$.
Now we repeat the steps we did for the DC-Hamiltonian. Assuming that
$2c^2\bm{1}+\bm{\mathcal{R}}^{(0)}$ is invertible and that the series
\begin{equation}
\bm{\tilde{\mathcal{R}}}(\tilde{x}) = \Sum{k=0}{\infty}
\underbrace{\left.\frac{1}{k!}\partiell{^k\bm{\tilde{\mathcal{R}}}}{\tilde{x}^k}\right|_{\tilde{x}=0}}_{\bm{\tilde{\mathcal{R}}}^{(k)}}
\tilde{x}^k
\end{equation}
converges, we can expand the inverse of the expression in parenthesis
in eq. (\ref{eq: gen_2c_alphavar}) and the transformation matrix
$\bm{\tilde{\mathcal{R}}}$ in $\tilde{x}$:
\begin{equation}
\begin{aligned}
\hat{\tilde{H}}_\mathrm{2c}&= \underbrace{\spinmom
c^2\bm{\omega}\bm{\tilde{\mathcal{R}}}\spinmom +
\hat{V}_{\mathrm{C}}}_{\hat{H}_\mathrm{2c}} \\
&+ \underbrace{\spinmom
c^2\bm{\omega}
\Sum{k=1}{\infty}
\brackets{\tilde{x}^k
\brackets{
\Sum{n=0}{k} \left(- \bm{\mathcal{R}}^{(0)} \bm{\omega}\right)^{k-n} \bm{\tilde{\mathcal{R}}}^{(n)} 
}
}
\spinmom
}_{\hat{H}_\mathrm{2c}(\delta\alpha)}\,,
\end{aligned}
\label{eq: 2c_perturbed}
\end{equation}
where we introduced the zeroth order transformation matrix operator
$c^2\bm{\omega}=\parantheses{2\times\bm{1}+\alpha^2\bm{\mathcal{R}}^{(0)}}^{-1}$

In case of one-particle regular expansions $\bm{\mathcal{R}}^{(0)}$
reduces to $-\hat{V}\times\bm{1}$ and $\bm{\tilde{\mathcal{R}}}$ is of
the form $\Sum{k=0}{n}\brackets{\frac{-E}{2c^2-\hat{V}}}^k\times\bm{1}$ for order
$n$ of the regular approximation. We will focus on zeroth order
(ZORA) using van W\"ullen's model potential $\tilde{V}$
\cite{wullen:1998}, such that $\bm{\mathcal{R}}^{(0)}$
reduces to the scalar potential $-\tilde{V}$ and
the inverse matrix $\bm{\omega}$ reduces to the scalar
$\omega=\parantheses{2c^2-\tilde{V}}^{-1}$. We obtain for the first
and second order perturbations in $\tilde{x}$
\begin{align}
\hat{H}^{(1)}_\mathrm{ZORA}&= \spinmom c^2\omega^2 \tilde{V} \spinmom 
\label{eq: zora_1order}\\
\hat{H}^{(2)}_\mathrm{ZORA}&= \spinmom c^2 \omega^3 \tilde{V}^2
\spinmom\,.
\label{eq: zora_2order}
\end{align}

These two Hamiltonians are then used in the same way as in eq.~\ref{eq: first_order_alpha_var} and
\ref{eq: second_order_alpha_var}, but in a two-component framework.

The Hamiltonians \ref{eq: zora_1order} and \ref{eq: zora_2order} can 
alternatively be obtained via the following route:
The perturbed one-particle Dirac equation can be written in $2\times2$ block form as
\begin{widetext}
\begin{equation}
\begin{pmatrix}
\hat{V}\bm{1} & c\vec{\pauli}\cdot\momop + \tilde{x} \bm{H}^{(1)}_\mathrm{LS} + \tilde{x}^2 \bm{H}^{(2)}_\mathrm{LS} + \mathcal{O}(\tilde{x}^3)\\
c\vec{\pauli}\cdot\momop + \tilde{x} \bm{H}^{(1)}_\mathrm{SL} + \tilde{x}^2 \bm{H}^{(2)}_\mathrm{SL} + \mathcal{O}(\tilde{x}^3) & \hat{V}\bm{1} - 2 c^2 +  \tilde{x} \bm{H}^{(1)}_\mathrm{SS} + \tilde{x}^2 \bm{H}^{(2)}_\mathrm{SS} + \mathcal{O}(\tilde{x}^3)\\
\end{pmatrix}
\Psi_{\mathrm{4c},i} = \epsilon_i  \Psi_{\mathrm{4c},i}
\label{eq: 4cmatformexpanded}
\end{equation}
Here we defined
$\bm{H}^{(1)}_\mathrm{LS}=\bm{H}^{(1)}_\mathrm{SL}=-\frac{c}{2}\vec{\pauli}\cdot\momop$,
$\bm{H}^{(1)}_\mathrm{SS}=2c^2\bm{1}$,
$\bm{H}^{(2)}_\mathrm{LS}=\bm{H}^{(2)}_\mathrm{SL}=\frac{3c}{8}\vec{\pauli}\cdot\momop$,
and $\bm{H}^{(2)}_\mathrm{SS}=-2c^2\bm{1}$.
The corresponding ZORA Hamiltonian is obtained by eliminating from eq.~\ref{eq: 4cmatformexpanded} the lower component of $\Psi_{\mathrm{4c},i}$ and approximating it subsequently by
$\psi^\mathrm{S}\rightarrow c\omega\spinmom\psi_\mathrm{ZORA}$. Retaining terms which are up to second order in $\tilde{x}$ we receive:
\begin{equation}
\begin{aligned}
\hat{H}_\mathrm{ZORA} &= \hat{H}_\mathrm{ZORA}^{(0)} \\
&+ \parantheses{ \tilde{x} \bm{H}^{(1)}_\mathrm{LS} + \tilde{x}^2 \bm{H}^{(2)}_\mathrm{LS}} \omega c\vec{\pauli}\cdot\momop \\
&+ c\vec{\pauli}\cdot\momop \omega \parantheses{ \tilde{x} \bm{H}^{(1)}_\mathrm{SL} + \tilde{x}^2 \bm{H}^{(2)}_\mathrm{SL}} \\
&+ \parantheses{ \tilde{x} \bm{H}^{(1)}_\mathrm{LS} } \omega^2 \parantheses{ \tilde{x} \bm{H}^{(1)}_\mathrm{SL}} \\
&+ c\vec{\pauli}\cdot\momop \omega \parantheses{ \tilde{x} \bm{H}^{(1)}_\mathrm{SS} + \tilde{x}^2 \bm{H}^{(2)}_\mathrm{SS} } \omega c\vec{\pauli}\cdot\momop  \\
&+ c\vec{\pauli}\cdot\momop \omega \parantheses{ \tilde{x} \bm{H}^{(1)}_\mathrm{SS} } \omega \parantheses{ \tilde{x} \bm{H}^{(1)}_\mathrm{SS} } \omega c\vec{\pauli}\cdot\momop  \\
\end{aligned}
\end{equation}
\end{widetext}
The resulting first and second order Hamiltonians in the ZORA picture are:
\begin{align}
\hat{H}^{(1)}_\mathrm{D,ZORA} &= -\spinmom c^2\omega\spinmom +
2c^4 \spinmom\omega^2\spinmom \\ 
\hat{H}^{(2)}_\mathrm{D,ZORA} &= \frac{3}{4}\spinmom c^2\omega\spinmom -
2c^4 \spinmom\omega^2\spinmom \nonumber\\
&+\frac{1}{4}\spinmom c^2\omega \spinmom
-2c^4 \spinmom\omega^2\spinmom \nonumber\\
&+ 4c^6 \spinmom\omega^3\spinmom\,,
\end{align}
where the last three terms of $\hat{H}^{(2)}_\mathrm{D,ZORA}$ are
diamagnetic-like contributions that stem from products of the large and small component blocks of
$\hat{H}^{(1)}_\mathrm{D,ZORA}$, which are of second order in $\tilde{x}$
in the transformation from the DC-Hamiltonian to ZORA, as is shown above.

This is in
agreement with eq. (\ref{eq: zora_1order}), as by using
$\omega=\frac{1}{2c^2}(1+\tilde{V}\omega)$ we receive
\begin{align}
\hat{H}^{(1)}_\mathrm{D,ZORA} &= c^2
\spinmom\omega\parantheses{2c^2\omega-1}\spinmom \nonumber\\
&= \spinmom
c^2\omega^2\tilde{V}\spinmom \label{eq:dczora1} \\
\hat{H}^{(2)}_\mathrm{D,ZORA} &= c^2
\spinmom\omega\parantheses{1-4c^2\omega+4c^4\omega^2}\spinmom
\nonumber\\
&= c^2
\spinmom\omega\parantheses{1-2c^2\omega}^2\spinmom
\nonumber\\
&= \spinmom
c^2\omega^3\tilde{V}^2\spinmom 
\,.
\end{align}  

Finally, it can be easily seen that the chosen expansion in $\tilde{x}$ is
equivalent to the Pauli-like expansion of the Dirac equation. For higher orders
in $\tilde{x}$ the present scheme could be related to the linear response with
elimination of small components (LRESC) \cite{aucar:2018}, which is designed to
capture all corrections of a given order in $\tilde{x}$ including those not
included in the DC picture such as the Breit interaction or virtual pair
creation. Using eq.  (\ref{eq: 2c_perturbed}) we can relate different orders of
the Pauli Hamiltonian to different orders in $\tilde{x}$: Within the Pauli-like
expansion $\bm{\mathcal{R}}^{(0)}$ vanishes and in $k$th order we have
\begin{equation}
\bm{\tilde{\mathcal{R}}}^{(k)}_\mathrm{Pauli}=\parantheses{\frac{V-E}{2c^2}}^k
\end{equation}
yielding the perturbing operators
\begin{align} 
\hat{H}^{(1)}_\mathrm{Pauli}&=\spinmom \frac{V-E}{4c^2} \spinmom\\
\hat{H}^{(2)}_\mathrm{Pauli}&=\spinmom
\frac{\parantheses{V-E}^2}{8c^4} \spinmom\,, 
\end{align} 
which are, as expected, just the relativistic corrections of different order in the
Pauli expansion.

\FloatBarrier
\section{Computational Details}%
Quasi-relativistic two-component calculations within the zeroth order regular
approximation (ZORA) were performed with a modified version
\cite{berger:2005,nahrwold:09,isaev:2012,gaul:2020,bruck:2023,colombojofre:2022,zulch:2022}
of a two-component program \cite{wullen:2010} based on Turbomole \cite{ahlrichs:1989} on the level of complex generalized
Hartree--Fock (cGHF). The gauge dependence of the ZORA Hamiltonian was
alleviated by a model potential as suggested by van W\"ullen \cite{wullen:1998}
which was applied with additional damping \cite{liu:1999}. 

Orbitals for excited Slater determinants were obtained by choosing the occupation numbers
in the SCF calculations to achieve maximum overlap with the
determinant of the previous cycle (maximum overlap method, MOM) as
described e.g.\ in Ref.~\cite{gilbert:2008}. In case that a change in the differential
density compared to the previous cycle was above $10^{-3}/a_0^{-3}$
the occupation numbers were chosen such that the maximum overlap of
the determinant with respect to the initial guess is achieved (initial
guess MOM, IMOM) \cite{gilbert:2008,barca:2018}. The initial guess was
generated by reoccupation of the cGHF orbitals (occupied and virtual)
obtained for the cGHF determinant of lowest energy. 

Expectation values of operators (\ref{eq: zora_1order}) and (\ref{eq:
zora_2order}) were computed via the toolbox approach outlined in
Ref.~\cite{gaul:2020}. Within the quasi-relativistic ZORA framework, the second order
expression in eq. (\ref{eq: second_order_alpha_var}) was computed at
the level of coupled perturbed HF as outlined in Refs.
\cite{gaul:2020,bruck:2023,colombojofre:2022}.

Transition electric dipole moments $\vec{\mu}^{\mathrm{f}\leftarrow\mathrm{i}}$ were computed using the
independently obtained cGHF determinants of the different states using
L{\"o}wdin rules \cite{Lowdin:55} for single-particle operator
transition matrix elements between non-orthogonal single-determinantal
initial (i) wave function $\Phi_\mathrm{i}$ and final (f) wave
function $\Phi_\mathrm{f}$ (see also Ref.~\cite{klues:2016}):

\begin{align} 
\begin{split} 
\vec{\mu}^{\mathrm{f}\leftarrow\mathrm{i}} 
  &= \Braket{\Phi_\mathrm{f} | \sum_k q_k \vec{r}_k | \Phi_\mathrm{i}} \\
  &= \sum_{ij} \Braket{\psi_{\mathrm{f},i} | -e\vec{r} |
    \psi_{\mathrm{i},j}} \text{adj}\left( \bm{S} \right)_{ij}
  + \sum_A Z_A e\vec{r}_A \det{\bm{S}} \,. \label{eq:loewdin}
\end{split} 
\end{align} 
Here, the index $k$ runs over all electrons and nuclei with charges $q_k$, whereas $A$ runs over all nuclei with nuclear charge number $Z_A$, with $e$ denoting the elementary charge. $\bm{S}$ is the overlap matrix
between the molecular spinors $\psi_i$ and $\psi_j$ forming the final and the initial state
Slater determinant, respectively, with corresponding matrix elements
$S_{ij}=\Braket{\psi_{\mathrm{f},i}|\psi_{\mathrm{i},j}}$. The symbol $\det{\bm{S}}$ 
denotes the determinant of the matrix $\bm{S}$, whereas $\text{adj}\left( \bm{S} \right)$ denotes the adjugate 
of $\bm{S}$. 
Wavefunctions were optimized until a change in energy lower than
$10^{-9}~E_{\text{h}}$ and in relative spin-orbit coupling energy contribution lower
than $10^{-11}$ was achieved. Optimizations of bond lengths were
performed until a change in energy of $10^{-6}~E_{\text{h}}$ or
lower was reached. Coupled perturbed HF equations were solved iteratively
until we obtained a norm of the residuum of $10^{-6}$ or lower (see Ref.~\cite{bruck:2023} for 
details of the implementation).

Relativistic four component calculations employing the DC-Hamiltonian were
performed with the quantum chemistry program package Dirac \cite{dirac19}. The
various states computed on the level of average-of-configuration (AOC)
Dirac--Hartree--Fock (DHF) were obtained in a MOM kind of approach while
treating the open-shell systems within the AOC framework.
Four component relativistic Fock-space coupled-cluster calculations with singles
and doubles cluster amplitudes (FSCCSD) were performed with the ground state
wavefunction of the closed shell \ce{CeF^3+} system as reference and attaching
one additional electron (0-1 sector). For FSCCSD calculations an active space with seven
electrons of F ($2s^2 2p^5$) and 19 electrons of Ce ($4d^{10} 5s^2 5p^6 4f^1$)
and virtual spinors up to an energy of \SI{50}{\hartree} was
employed.  \ce{PaF^3+} was treated as described in Ref.~\cite{zulch:2022}. 
AOC-DHF spinors were optimized until a change in the orbital gradient
\SI{e-8}{} or lower was reached. 

In all calculations with \ce{CeF^2+} Dyall's core-valence triple-$\zeta$
(Dyall-cv3z) basis set was employed for Ce and F
\cite{gomes:2010,dyall:2002,dyall:2006}. In Ref.~\cite{simpson:2025} we
employed a similar basis set in the two-component ZORA-cGHF computations
of \ce{CeF^2+}, but therein with additional steep s and p functions. Those additional
steep functions weakly impact on the electronic properties 
computed in our present work. Consequently, also
potential energy curves with similar shapes and 
similar spectroscopic constants derived from these potential energy curves
(see Tab.~\ref{tab:props}) result. For completeness, we provide the present
raw data of the potential energy curves (computed at
$c=c_0=137.0359895~a_0E_\mathrm{h}\hbar^{-1}$) in the Supplemental Material
\cite{ref:si} and display therein also the corresponding potential energy
curves.

For the computation of \ce{PaF^3+} we followed Ref.~\cite{zulch:2022} and used
in two-component calculations
with \ce{PaF^3+} a basis set consisting of 37s, 34p, 14d and 9f
uncontracted Gaussian functions for Pa. The basis set is
constructed as an even-tempered series as $\alpha_i=a/b^{i-1}$;
$i=1,...,N$, with the exponential coefficients $\alpha_i$ and $N$ as
the number of functions, with $b=2$ for s and p functions and
$b=(5/2)^{1/25} \times 10^{2/5}\approx 2.6$ for d and f functions.
The corresponding largest exponent coefficients $a$ of the subsets are
$2\times10^9~a_0^{-2}$ (s), $5\times 10^8~a_0^{-2}$ (p),
$13300.758~a_0^{-2}$ (d) and $751.8368350~a_0^{-2}$ (f). For the F
atom in these ZORA-cGHF computations a decontracted atomic natural orbital
basis set of double-$\zeta$
quality augmented with polarization valence basis functions
(ANO-RCC-VDZP) \cite{roos:2004} was employed.
We note that, in contrast to Ref.~\cite{zulch:2022},
a wavefunction fulfilling tighter convergence criteria in the spin-orbit coupling
contribution was required for the subsequent response calculations.

In all two- and four-component calculations a normalized spherical
Gaussian nuclear density distribution $\rho_A \left( \vec{r}
\right) = \frac{\zeta_A^{3/2}}{\pi ^{3/2}} \text{e}^{-\zeta_A \left|
\vec{r} - \vec{r}_A \right| ^2}$ with corresponding charge distribution
$Z_A e \rho_A \left( \vec{r} \right)$ was employed with $\zeta_A =
\frac{3}{2 r^2 _{\text{nuc},A}}$, and the root-mean-square radius
$r_{\text{nuc},A}$ chosen as suggested by Visscher and Dyall
\cite{visscher:1997} with the isotopes $^{140}$Ce, $^{231}$Pa and
$^{19}$F. 

To obtain for comparison to our perturbative approach the sensitivity to variations of $\alpha$ also numerically, we altered the
speed of light $c$, which is (in atomic units) $c/(a_0 E_\mathrm{h} \hbar^{-1})=1/\alpha=137.0359895$, to take the
values $125.0359895+n$ where $n$ are integer values from $0$ to $26$. To
densify the region around $c/(a_0\,E_\mathrm{h}\,\hbar^{-1})=137.0359895$ we further included the values
136.6359895, 136.7359895, 136.8359895, 136.9359895, 137.1359895, 137.2359895,
137.3359895, 137.4359895 on the level of ZORA-cGHF and AOC-DHF. For the FSCCSD
calculations we varied $c/(a_0\,E_\mathrm{h}\,\hbar^{-1})$ to take the values 125.0359895, 127.0359895,
129.0359895, 131.0359895, 133.0359895, 135.0359895, 137.0359895, 140.0359895,
143.0359895, 146.0359895, 149.0359895, 152.0359895. The sensitivity parameters
are then obtained via a fit to a 4th order polynomial. As fitting procedure  a nonlinear
least squares Marquardt--Levenberg algorithm as implemented in \textsc{Gnuplot}
\cite{gnuplot} was employed. We report numerical data and raw data from the outputs of the
calculations in the Supplemental Material \cite{ref:si}.

 From the potential
energy curves electronic excitation wavenumbers $\tilde{\nu}$, equilibrium
bond lengths $r_\mathrm{e}$ and harmonic vibrational wavenumbers
$\tilde{\omega}_\mathrm{e}$ are obtained by fitting a harmonic oscillator potential around
the potential minimum (data points are listed in Supplemental Material \cite{ref:si}).

For a vibronic transition analysis vibrational corrections to the property
$q_{1}$ are computed by solving the vibrational Schr\"odinger equation in a
one-dimensional discrete variable representation (DVR) approach on an
equidistant grid \cite{meyer:1970}. The corresponding grid points for each state
are given in Tab.~\ref{tab:dvrPoints}. We interpolate $q_{1}$ as a polynomial
function of the bond length of order three for the electronic ground state and
of order five for electronic excited states as described e.g.\ in the
Supplemental Material of Ref.~\cite{udrescu:2021}. Natural radiative linewidths
$\Gamma_{iv}$ are approximated by summing the Einstein coefficients
$A_{ju}$ of all decay channels from the considered excited state (stated here in SI units)
\begin{align}
\begin{split}
   \Gamma_{iv} &= \sum_{ju} \theta\left( E_{iv} - E_{ju} \right) A_{ju,iv} \\
   &= \sum_{ju} \theta\left( E_{iv} - E_{ju} \right)\frac{8\pi^2\tilde{\nu}_{ju,iv}^3}{3\epsilon_0\hbar} \left|
   \vec{\mu}^2_{ju,iv} \right|^2
\label{eq:linewidths}
\end{split}
\end{align}
with the electric constant  $\epsilon_0$ and the reduced Planck constant $\hbar$.
The transitions are defined through the vibrational $(u,v)$ and electronic $(j,i)$
state.  Here, we used a left-continuous Heaviside-function $\theta\left( E_{iv} -
E_{ju} \right)$ that yields $0$, when the energetic difference $E_{iv} - E_{ju}$
between the vibronic levels is zero or lower, and $1$ otherwise. The vibronic
electric transition dipole moment $\vec{\mu}_{ju,iv}$ (when $i$
and $j$ differ) is estimated by multiplying the Franck--Condon factor of the
transition $f^{(ju,iv)}$ with the squared electronic transition dipole moment
$\vec{\mu}^{j\leftarrow i}$ computed at the equilibrium
structure of the electronic ground state. The matrix elements of
the vibrational electric transition dipole moment (when $i$ and $j$ are the same)
are computed within the DVR ansatz using the computed bond length dependence of
the total electric dipole moment in the given electronic state.

\begin{table}
\centering
\caption{Grid point parameters for the DVR for the different electronic states.
The evenly spaced grid starts from the internuclear distance
$r_\mathrm{lowest}$ and is increased in increments of $r_\mathrm{step}$ to
reach the total number of grid points $n$.}
\label{tab:dvrPoints}
\begin{threeparttable}
\begin{tabular}{
c
S[table-format=1.2]
S[table-format=1.5]
S[table-format=4.0]
c
S[table-format=1.2]
S[table-format=1.5]
S[table-format=4.0]
}
\toprule
  \multirow{2}{*}{State}
  & \multicolumn{3}{c}{\ce{CeF^2+}}
& \phantom{aa} & \multicolumn{3}{c}{\ce{PaF^3+}}\\

& {$r_\mathrm{lowest} / \si{\angstrom}$}
& {$r_\mathrm{step} / \si{\angstrom}$}
& {$n$}
&
& {$r_\mathrm{lowest} / \si{\angstrom}$}
& {$r_\mathrm{step} / \si{\pico\meter}$}
& {$n$}\\
\midrule
(X)5/2 & 1.34 &  .001  & 2400  & & 1.3  &   .175 & 1000  \\
(1)3/2 & 1.34 &  .001  & 2400  & & 0.95 &   .100 & 1900  \\
(1)1/2 & 1.34 &  .001  & 2400  & & 0.95 &   .100 & 2000  \\
(1)7/2 & 1.34 &  .001  & 2400  & & 0.95 &   .100 & 2090  \\
(1)5/2 & 1.34 &  .001  & 2400  & & 0.95 &   .100 & 2090  \\
(2)3/2 & 1.34 &  .001  & 2400  & & 0.95 &   .100 & 2090  \\
(2)1/2 & 1.34 &  .001  & 2400  & & 1.15 &   .100 & 1850  \\
(3)3/2 & 1.34 &  .001  & 1910  & & 1.40 &   .100 & 950   \\
\bottomrule
\end{tabular}
\end{threeparttable}
\end{table}

\begin{figure}
  \includegraphics[width=0.45\textwidth]{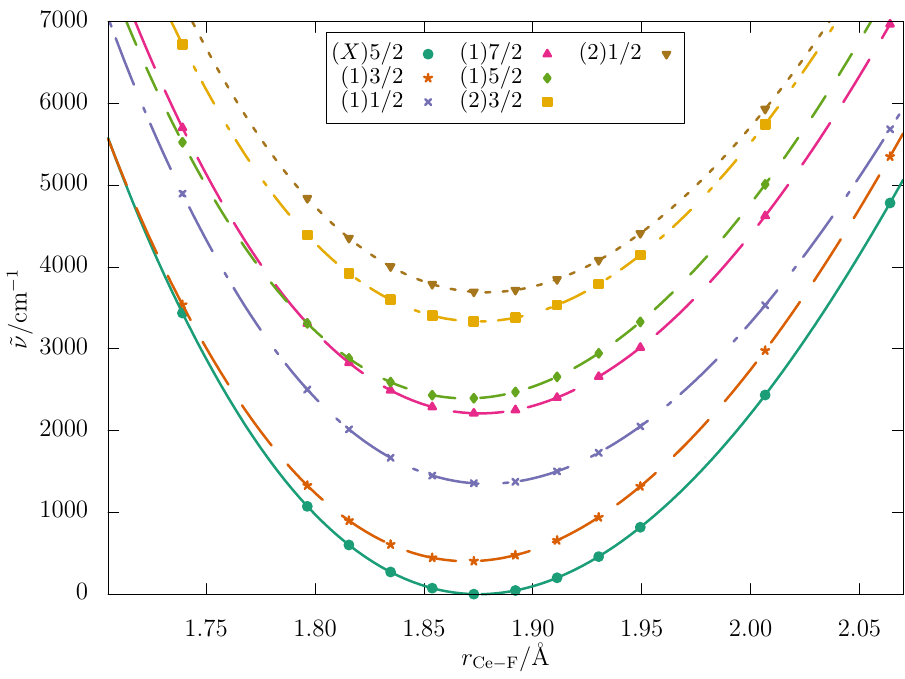}
  \caption{Potential energy curves for the seven energetically lowest
  electronic states relative to the energetic minimum of the electronic ground
  state of \ce{CeF^2+} computed on the level of DC-FSCCSD.}
  \label{fig:pot2}
\end{figure}

\begin{table*}
\begin{threeparttable}
\centering
\caption{Equilibrium bond lengths $r_\mathrm{e}$, vertical excitation
wavenumber $\tilde{\nu}$ from the electronic ground state minimum  and harmonic
vibrational wavenumber $\tilde{\omega}_\mathrm{e}$ of \ce{CeF^2+} as obtained
from potential energy curves on the level of ZORA-cGHF, AOC-DHF and FSCCSD.  We
received for \ce{CeF^2+} on the ZORA-cGHF level values that are within the
reported number of digits identical to those computed in
Ref.~\cite{simpson:2025} with a basis set that was augmented by additional
steep s and p functions (see computational details for different basis sets
employed). Where available data are compared to \ce{PaF^3+} from Ref.~\cite{zulch:2022}. 
We note that the basis sets employed for \ce{CeF^2+} and
\ce{PaF^3+} differ in their quality.}
\label{tab:props}
\begin{tabular}{ 
ll 
S[table-format=2.3,round-mode=figures,round-precision=3]
S[table-format=7.3,round-mode=figures,round-precision=3]
S[table-format=4.2,round-mode=figures,round-precision=3]
c
S[table-format=2.3,round-mode=figures,round-precision=3]
S[table-format=7.3,round-mode=figures,round-precision=3]
c
l
}
\toprule
  \multirow{2}{*}{State}   
& \multirow{2}{*}{Method} 
& \multicolumn{3}{c}{\ce{CeF^2+}} & & \multicolumn{3}{c}{\ce{PaF^3+ \cite{zulch:2022}}}\\
&     
& {$r_\mathrm{e} / \si{\angstrom}$} 
& {$\tilde{\nu} / \si{cm^{-1}}$} 
& {$\tilde{\omega}_\mathrm{e} / \si{cm^{-1}}$}  
&
& {$r_\mathrm{e} / \si{\angstrom}$} 
& {$\tilde{\nu} / \si{cm^{-1}}$} 
& {$\tilde{\omega}_\mathrm{e} / \si{cm^{-1}}$} \\ 
\midrule
\multirow{2}{*}{(X)5/2} &  ZORA-cGHF &  1.912 & {\textemdash} & 722  & & 1.87 & {\textemdash} & 846~{\tnote{(1)}}\\
                        &  AOC-DHF   &  1.912 & {\textemdash} & 722  & & \\
                        &  FSCCSD    &  1.875 & {\textemdash} & 803  & & 1.85 & {\textemdash} & 859~\vphantom{{\tnote{(1)}}} \\
\\
\multirow{2}{*}{(1)3/2} &  ZORA-cGHF &  1.907 & 1050          & 718  & & 1.86  & 1250         & 842~{\tnote{(1)}} \\
                        &  AOC-DHF   &  1.907 & 1217          & 716  & & \\
                        &  FSCCSD    &  1.875 & 404           & 803  & & 1.84  & 658          & 844~\vphantom{{\tnote{(1)}}} \\
\\
\multirow{2}{*}{(1)1/2} &  ZORA-cGHF &  1.914 & 1460          & 716  & & 1.87  & 3020         & 838~{\tnote{(1)}} \\
                        &  AOC-DHF   &  1.914 & 1856          & 716  & & \\
                        &  FSCCSD    &  1.880 & 1356          & 788  & & 1.86  & 3060         & 861~\vphantom{{\tnote{(1)}}} \\
\\
\multirow{2}{*}{(1)7/2} &  ZORA-cGHF &  1.912 & 2304          & 723  & & 1.87  & 5520         & 846~{\tnote{(1)}} \\
                        &  AOC-DHF   &  1.912 & 2125          & 722  & & \\
                        &  FSCCSD    &  1.876 & 2212          & 804  & & 1.85  & 5540         & 861~\vphantom{{\tnote{(1)}}} \\
\\
\multirow{2}{*}{(1)5/2} &  ZORA-cGHF &  1.906 & 3306          & 720  & & 1.86  & 6440         & 845~{\tnote{(1)}} \\
                        &  AOC-DHF   &  1.906 & 3242          & 715  & & \\
                        &  FSCCSD    &  1.869 & 2396          & 801  & & 1.84  & 5790         & 846~\vphantom{{\tnote{(1)}}} \\
\\
\multirow{2}{*}{(2)3/2} &  ZORA-cGHF &  1.911 & 3765          & 718  & & 1.87  & 8000         & 841~{\tnote{(1)}} \\
                        &  AOC-DHF   &  1.911 & 3824          & 717  & & \\
                        &  FSCCSD    &  1.875 & 3336          & 798  & & 1.85  & 7810         & 853~\vphantom{{\tnote{(1)}}}\\
\\
\multirow{2}{*}{(2)1/2} &  ZORA-cGHF &  1.914 & 3927          & 716  & & 1.87  & 8820         & 840~{\tnote{(1)}} \\
                        &  AOC-DHF   &  1.915 & 4232          & 718  & & \\
                        &  FSCCSD    &  1.879 & 3698          & 791  & & 1.86  & 8940         & 858~\vphantom{{\tnote{(1)}}} \\
\\
\multirow{2}{*}{(3)3/2} &  ZORA-cGHF &  1.847 & 31903         & 774  & & 1.83  & 30000        & 888~{\tnote{(1)}} \\
                        &  AOC-DHF   &  1.847 & 31294         & 755  & & \\
                        &  FSCCSD    &  1.879 & 27872         & 791  & & 1.83  & 29200        & 882~\vphantom{{\tnote{(1)}}} \\
\bottomrule
\end{tabular}
\begin{tablenotes}
  \item[(1)] The harmonic vibrational wavenumbers for \ce{PaF^3+} on the level of ZORA-cGHF are obtained
  herein by fitting a third order polynomial to the numerically calculated
  potential energy curve from Ref.~\cite{zulch:2022}, whereas
  in Ref.~\cite{zulch:2022} a second order polynomial
  was used. This causes small numerical differences in the reported values. Both fit procedures only use data 
  points around the respective equilibrium bond lengths.
\end{tablenotes}
\end{threeparttable}
\end{table*}

\section{Results}%
\subsection{Electronic states of \ce{CeF^2+} and \ce{PaF^3+}} The electronic
structure of \ce{PaF^3+} was discussed in detail in Ref.~\cite{zulch:2022}. For
completeness we present herein again spectroscopic properties of the eight
energetically lowest electronic states in \ce{PaF^3+} and compare them to the
electronic states of \ce{CeF^2+} in Table~\ref{tab:props}. For \ce{CeF^2+} we emphasise, that
the basis set employed herein in the two-component ZORA-cGHF calculations is
similar to the basis set employed in Ref.~\cite{simpson:2025}, but omits the additional steep s and p functions. This similarity of basis sets for \ce{CeF^2+}
results in nearly identical potential energy shapes obtained on the level of ZORA-cGHF compared
to Ref.~\cite{simpson:2025} (see computational
details). In the present paper we additionally report potential energy curves
computed on the level of FSCCSD of \ce{CeF^2+} (Fig.~\ref{fig:pot2}). Derived from the
potential energy curves, we obtain relevant spectroscopic properties that are
compared in Tab.~\ref{tab:props}. 

As previously analysed in Ref.~\cite{simpson:2025}, \ce{CeF^2+} has a similar electronic structure as
\ce{PaF^3+} and only slightly shorter bond lengths. Furthermore, in
\ce{CeF^2+} the energetically lowest seven electronic states are even closer to each
other in energy than in
\ce{PaF^3+}. The agreement between ZORA-cGHF and FSCCSD is similar as
in \ce{PaF^3+}: ZORA-cGHF gives in general reasonable results, with
all vertical excitation wavenumbers appearing slightly
overestimated by the mean field approach. This effect is larger for the lower states computed with AOC-DHF
due to lacking spin-polarisation in the AOC-DHF approach.

\begin{table*}
\begin{threeparttable}
\centering
\caption{Analytically (an.) and numerically (num.) determined results for the
sensitivity parameters $\Delta q_{1}$ and $2\Delta q_{2}$ for \ce{CeF^2+}
computed on the level of ZORA-cGHF and AOC-DHF. 
$\Delta q_i$ are computed with the electronic ground state as reference as $\Delta q_i = q_i\left(
\mathrm{excited\ state} \right) - q_i\left( \mathrm{ground\ state} \right)$.
The relative deviation between
ZORA-cGHF and AOC-DHF was computed as $\mathrm{dev._{4c,2c}}=1-\frac{\Delta
q_{\mathrm{ZORA-cGHF}}}{\Delta q_{\mathrm{AOC-DHF}}}$, where in case of $\Delta
q_1$ the analytical values were used and in case of $2\Delta q_2$ the analytical
value from ZORA-cGHF and the numerical value from AOC-DHF. For the $(3)3/2$
electronic state only analytical results are computed. For each method and each
electronic state the corresponding equilibrium bond length was used as
internuclear distance (see Supplemental Material). }
\label{tab:cghf_dhf}
\begin{tabular}{
ll
S[table-format=-5.3]
S[table-format=-5.6]
r
S[table-format=-5.3]
S[table-format=-4.6]
r
}
\toprule
  \multirow{2}{*}{State}
& \multirow{2}{*}{Method}
& \multicolumn{3}{c}{$\Delta q_{1} / \si{cm^{-1}}$}
& \multicolumn{3}{c}{$2\Delta q_{2} / \si{cm^{-1}}$} \\
&
& {$\mathrm{an.} $}
& {$\mathrm{num.}$}
& {$\mathrm{dev._{4c,2c}}/\%$}
& {$\mathrm{an.} $}
& {$\mathrm{num.}$}
& {$\mathrm{dev._{4c,2c}}/\%$} \\
\midrule
\multirow{2}{*}{(1)3/2} &  ZORA-cGHF & 123.3   & 123.3  & {\multirow{2}{*}{$ 7.3$}}  & -32           & -31  & {\multirow{2}{*}{$48.4$}} \\
                        &  AOC-DHF   & 133.0   & 133.0  & {                     }  & {\textemdash} & -61  & {                     } \\
\\
\multirow{2}{*}{(1)1/2} &  ZORA-cGHF & -222.9  & -222.9 & {\multirow{2}{*}{$-10.8$}} & -272          & -272 & {\multirow{2}{*}{$35.6$}} \\
                        &  AOC-DHF   & -201.3  & -201.3 & {                     }  & {\textemdash} & -423 & {                     } \\
\\
\multirow{2}{*}{(1)7/2} &  ZORA-cGHF & 2023.2  & 2023.3 & {\multirow{2}{*}{$-8.5$}}  & -727          & -727 & {\multirow{2}{*}{$-8.6$}} \\
                        &  AOC-DHF   & 1865.6  & 1865.6 & {                     }  & {\textemdash} & -669 & {                     } \\
\\
\multirow{2}{*}{(1)5/2} &  ZORA-cGHF & 2186.2  & 2186.3 & {\multirow{2}{*}{$-8.0$}}  & -734          & -734 & {\multirow{2}{*}{$-8.0$}} \\
                        &  AOC-DHF   & 2024.1  & 2024.1 & {                     }  & {\textemdash} & -679 & {                     } \\
\\
\multirow{2}{*}{(2)3/2} &  ZORA-cGHF & 2078.3  & 2078.3 & {\multirow{2}{*}{$-7.9$}}  & -867          & -867 & {\multirow{2}{*}{$-5.1$}} \\
                        &  AOC-DHF   & 1926.6  & 1926.6 & {                     }  & {\textemdash} & -825 & {                     } \\
\\
\multirow{2}{*}{(2)1/2} &  ZORA-cGHF & 1744.7  & 1744.8 & {\multirow{2}{*}{$-12.7$}} & -998          & -998 & {\multirow{2}{*}{$-4.8$}} \\
                        &  AOC-DHF   & 1547.8  & 1547.8 & {                     }  & {\textemdash} & -953 & {                     } \\
\multirow{2}{*}{(3)3/2} &  ZORA-cGHF & -23863  & {\textemdash} & {\multirow{2}{*}{$1.7$}} & -2434         & {\textemdash} & {\multirow{2}{*}{\textemdash}} \\
                        &  AOC-DHF   & -24283  & {\textemdash} & {                      }  & {\textemdash}   & {\textemdash} & {                     } \\
\bottomrule
\end{tabular}
\end{threeparttable}
\end{table*}

\begin{figure}
  \includegraphics[width=0.45\textwidth]{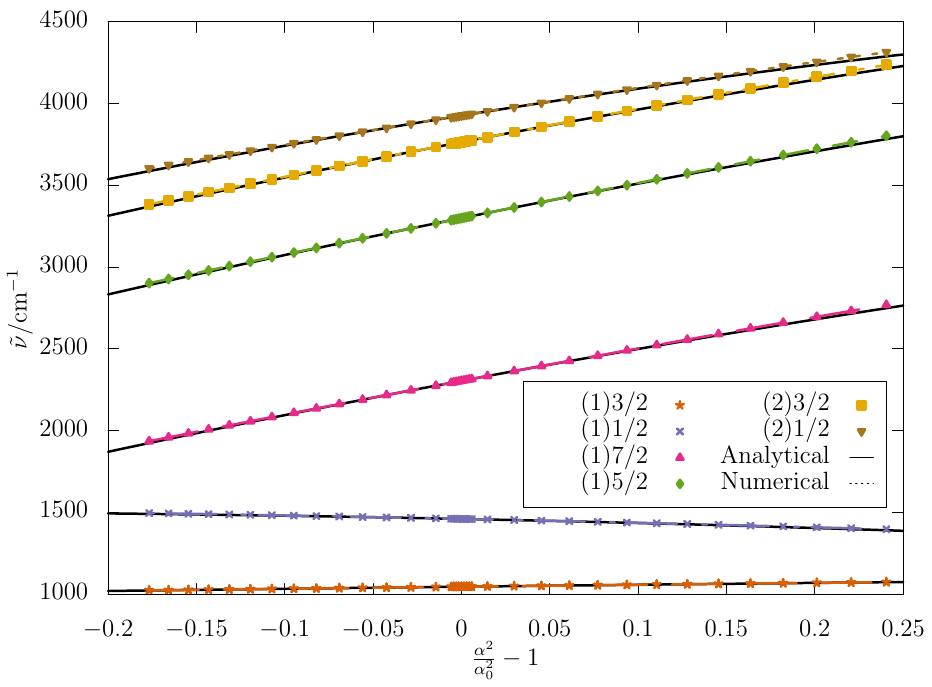}
  \caption{Adiabatic excitation energy of \ce{CeF^2+} as a function of the variation in
  $\alpha$ computed on the ZORA-cGHF level.
  The bond lengths for the different electronic states
  are listed in Tab.~\ref{tab:props} and were not changed when varying
  $\alpha$.
  Dots correspond to the values obtained when changing $\alpha$ numerically with the 
  corresponding 4th order polynomial fit shown as dashed lines. Solid 
  lines correspond to the analytically obtained second order expansion with $\Delta q_1$ and 
  $\Delta q_2$ values from Tab.~\ref{tab:cghf_dhf}. 
  }
   \label{fig:vdzp_all}
\end{figure}

\begin{figure}
  \includegraphics[width=0.45\textwidth]{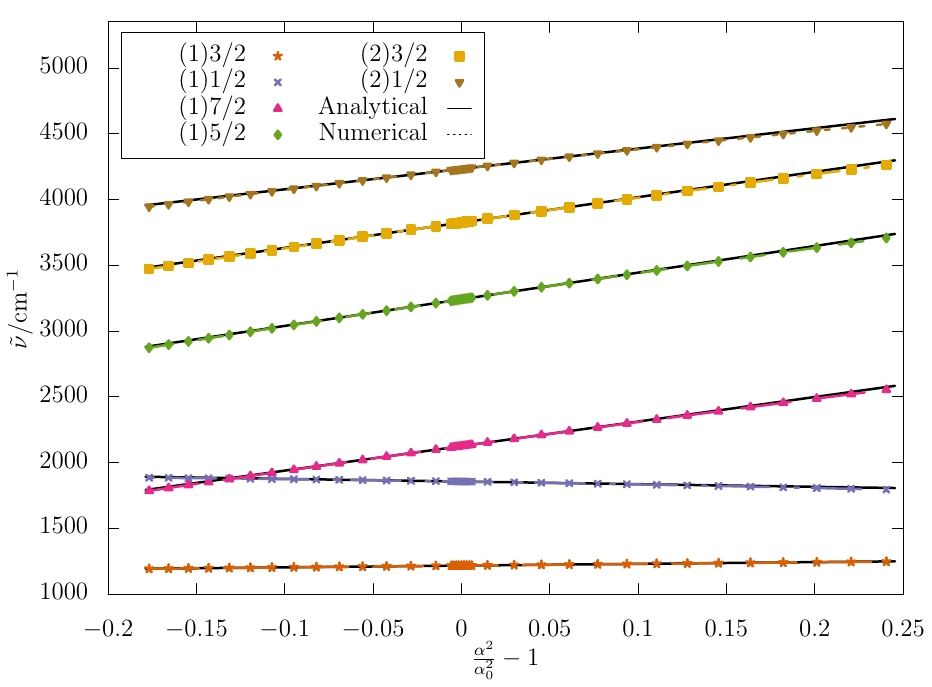}
  \caption{Adiabatic excitation energy of \ce{CeF^2+} as a function of the variation in
  $\alpha$ computed on the AOC-DHF level.
  The bond lengths for the different electronic states
  are listed in Tab.~\ref{tab:props} and were not changed when varying
  $\alpha$.
  Dots correspond to the values obtained when changing $\alpha$ numerically with the 
  corresponding 4th order polynomial fit shown as dashed lines. Solid 
  lines correspond to the analytically obtained second order expansion with $\Delta q_1$ and 
  $\Delta q_2$ values from Tab.~\ref{tab:cghf_dhf}. 
  }
   \label{fig:dhf_all}
\end{figure}

\begin{figure}[!htb]
 \centering
 \includegraphics[width=0.45\textwidth]{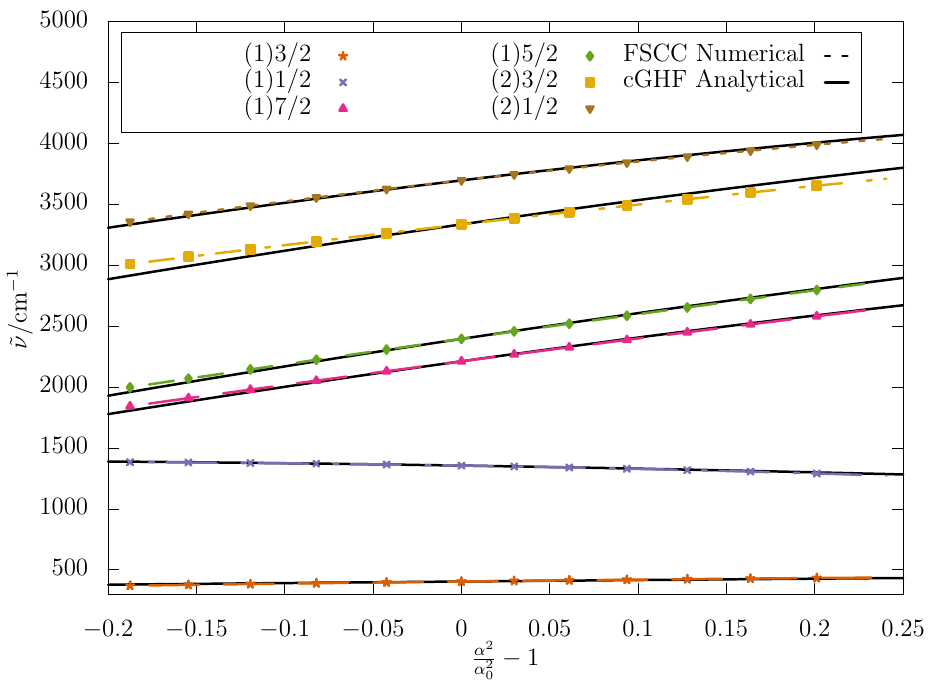}
  \caption{Adiabatic excitation energy of \ce{CeF^2+} as a function of the
  variation in $\alpha$ computed on the FSCCSD level. The bond lengths for the
  different electronic states are listed in Tab.~\ref{tab:props} and were not
  changed when varying $\alpha$.  Dots correspond to the values obtained when
  changing $\alpha$ numerically with the corresponding 4th order polynomial fit
  shown as dashed lines.  For comparison, curves with slopes and curvatures
  obtained analytically obtained on the ZORA-cGHF level are shown with their
  y-intercepts set to the corresponding FSCCSD values.
  }
 \label{fig:fscc_around0}
\end{figure}

\begin{table}
\resizebox{0.5\textwidth}{!}{%
\begin{threeparttable}
\caption{Numerical results for the sensitivity parameters $\Delta q_{1}$ and $2\Delta q_{2}$
for \ce{CeF^2+} computed on the level of FSCCSD compared to the results computed on the
level of AOC-DHF and ZORA-cGHF with the electronic ground state as reference.
$\Delta q_i$ are computed with the electronic ground state as reference as $\Delta q_i = q_i\left(
\mathrm{excited\ state} \right) - q_i\left( \mathrm{ground\ state} \right)$.
The relative deviations were computed as
$\mathrm{dev._\mathrm{method}}=1-\frac{\Delta q_{\mathrm{method}}}{\Delta q_{\mathrm{FSCCSD}}}$.
For ZORA-cGHF and AOC-DHF the equilibrium bond lengths for each respective
electronic state was used, while in FSCCSD the internuclear distance
was set for all electronic states to the AOC-DHF equilibrium value
of the electronic ground state (see Supplemental
Material \cite{ref:si}).}
\label{tab:fsccsd_comp}
\begin{tabular}{l 
S[table-format=5.2]
S[table-format=-4.1]
S[table-format=-2.1]
S[table-format=-2.1]
S[table-format=-2.1]
}
\toprule
{State} & 
{$\Delta q_{1,\mathrm{FSCCSD}}$ } &   
{$2\Delta q_{2,\mathrm{FSCCSD}}$ } &   
{$\Delta q_{1,\mathrm{DHF}}^\mathrm{dev.}/\%$ } & 
{$\Delta q_{1,\mathrm{cGHF} }^\mathrm{dev.}/\%$} & 
{$\Delta q_{2,\mathrm{cGHF} }^\mathrm{dev.}/\%$} \\ 
\midrule
(1)3/2  &  163.9      & -191        & 18.9  & 24.8  & 83.4  \\
(1)1/2  & -229.5      & -859        & 12.3  &  2.9  & 68.3  \\
(1)7/2  &  1907.8     & -668        &  2.2  & -6.1  & -8.8  \\
(1)5/2  &  2061.5     & -665        &  1.8  & -6.0  & -10.4 \\
(2)3/2  &  1671.2     & -781        & -15.3 & -24.4 & -11.0 \\
(2)1/2  &  1639.6     & -1821       &  5.6  & -6.4  & 45.2  \\
(3)3/2  &  -19114     & -4449       & -27.0 & -24.8 & -54.9 \\
\bottomrule
\end{tabular}
\end{threeparttable}
}
\end{table}

\subsection{Comparison of analytical vs. numerical computation of
$\alpha$-variations in \ce{CeF^2+}}

\begin{figure*}
   \centering
   \includegraphics[trim={0cm 0cm 0cm 0.3cm},clip,width=0.75\textwidth]{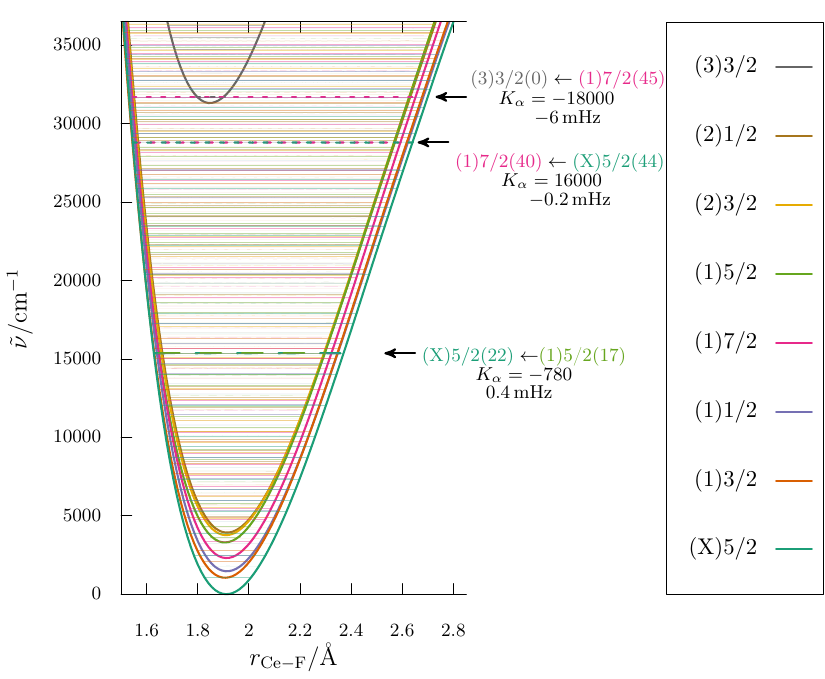}
   \vspace{-0.5cm}
   \includegraphics[trim={0cm 0cm 0cm 0.3cm},clip,width=0.75\textwidth]{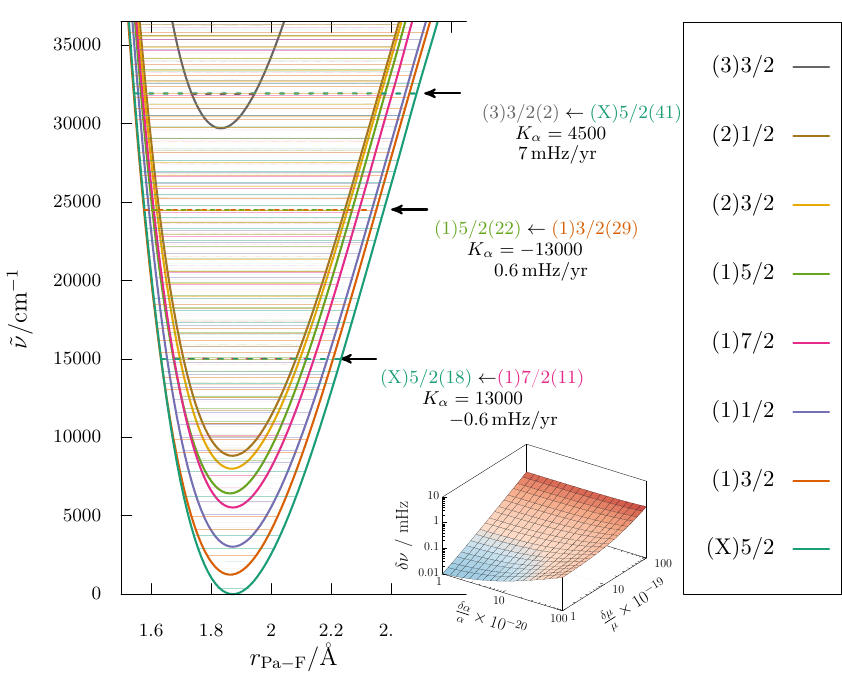}
   \caption{Born-Oppenheimer potential energy curves for \ce{CeF^2+} (top)
   and \ce{PaF^3+} (bottom, \cite{zulch:2022}) computed on the level 
   of ZORA-cGHF shown with the
   respective vibrational energy levels obtained herein with a DVR approach.
   For \ce{CeF^2+}, we emphasise the similarity to the potential
   energy curves reported in Ref.~\cite{simpson:2025} due to a similar 
   basis sets employed (see computational details).
   Exemplary states with pronounced predicted
   wavenumber shifts due to a variation of the fine-structure constant and
   the electron-proton mass-ratio are depicted. 
   The surface plot on the bottom
   shows the frequency shift of the transition
   $(1)7/2(11) \leftarrow (\mathrm{X})5/2(18)$ in \ce{PaF^3+} as function of 
   $\delta\alpha/\alpha$ and $\delta\mu/\mu$, where the $\alpha$-variation
   was computed using the respective equilibrium structures.}
   \label{fig:fancy_ce}
\end{figure*}

In this section we illustrate the validity of our analytical approach for the calculation of the
sensitivity parameters $q_1$ and $q_2$ by comparison to numerical variations of the
fine structure constant in AOC-DHF and ZORA-cGHF calculations. The
comparison between the numerical and analytical results is presented in
Tab.~\ref{tab:cghf_dhf} (the corresponding vertical excitation
energies as a
function of $\alpha$ are shown in Figures~\ref{fig:vdzp_all} to
\ref{fig:fscc_around0}). 
We report $\Delta q_1$ and $\Delta q_2$ as the difference with respect to the electronic ground
state as $\Delta q_i = q_i\left( \mathrm{excited\ state} \right) - q_i\left( \mathrm{ground\ state} \right)$
(for a discussion of the use of units in connection with a variation of dimensionless fundamental
constants, see, for example, Refs.~\cite{beloy:2010,beloy:2011a,kozlov:2019}).

All numerical results for $\Delta q_{1}$ are in excellent agreement with the
analytical calculations. The comparison between analytical and numerical approach
to the second order parameter $\Delta q_{2}$ is only done at the level of ZORA-cGHF as
in the program package DIRAC open-shell linear response calculations are not available.
The values of $\Delta q_{1}$ computed on the level of
ZORA-cGHF deviate from values computed on the level of AOC-DHF by $\lesssim
13\%$. This deviation is partially explainable by the different HF
approaches used in ZORA-cGHF and AOC-DHF calculations, which is also reflected
in the deviation of the excitation wavenumbers in Tab.~\ref{tab:props}.
Deviations are even larger for $q_{2}$, which may depend more strongly on the
choice of the relativistic Hamiltonian, as it may include
two-electron exchange contributions from spin-orbit coupling which we neglect
in our ZORA calculations. Moreover, the different HF method, i.e. AOC-HF and
cGHF, include correlation effects to some extend but in very different
ways. To gauge  errors of both HF approaches we performed more sophisticated
FSCCSD calculations of $q_{1}$ and $q_{2}$ using the numerical approach
(Tab.~\ref{tab:fsccsd_comp}).

FSCCSD results for $\Delta q_{1}$ are in reasonable agreement with ZORA-cGHF.
Deviations for $\Delta q_{2}$ are larger, which may in part be attributed to the stronger
dependence on two-electron exchange contributions to spin-orbit coupling.

\subsection{Enhancement of variations of fundamental constants in
vibronic transitions of \ce{CeF^2+} and \ce{PaF^3+}} 

Both molecular ions \ce{CeF^2+} and \ce{PaF^3+} possess several energetically close lying
electronic states. Consequently, a huge variety of accidentally nearly
degenerate vibronic states can be expected. Particular
pronounced shifts are expected in the region in which the vibrational levels of the energetically lowest seven potentials, which
are dominated by an unpaired f-like electron located at Ce or Pa,
are close to the vibrational levels of the seventh excited states, which is dominated by
an unpaired d-like electron located at Ce or Pa (see discussion in
Ref.~\cite{zulch:2022} and Fig.~\ref{fig:fancy_ce}). In an experiment,
nearly degenerate vibronic levels can be used \cite{demille:2008a}
to measure their
transition wavenumbers $\tilde{\nu}_{iv,jw}$
and the variation of the following fraction over time
\begin{equation}
\begin{aligned}
\frac{\delta\tilde{\nu}_{iv,jw}}{\tilde{\nu}_{iv,jw}}
&= 
K_{\alpha,iv,jw} \frac{\delta\alpha}{\alpha} + K_{\mu,iv,jw}\frac{\delta\mu}{\mu}\,,
\end{aligned}
\end{equation}
where 
\begin{align}
K_{\alpha,iv,jw} &=\frac{2(q_{1,iv}- q_{1,jw})}{\tilde{\nu}_{iv,jw}}
=\frac{2(\Delta q_{1,iv}- \Delta q_{1,jw})}{\tilde{\nu}_{iv,jw}}\,,\\
K_{\mu,iv,jw} &=\frac{\tilde{\nu}_{iv,jw}-\tilde{\nu}_{i,j}}
                {2\tilde{\nu}_{iv,jw}}\,,
\end{align}
where $\tilde{\nu}_{i,j}$ is the adiabatic electronic transition energy.
The importance of close lying states for large enhancements of a
hypothetical $\alpha$ variation is e.g. demonstrated with the literature known
atomic clock Yb \cite{jiang:2011,hinkley:2013,dzuba:2018a}. The estimated $\Delta q_1$
values for transitions from the Yb ground state to excited electronic states
are in the range of \SI{1000}{cm^{-1}} and \SI{-40000}{cm^{-1}}
\cite{safronova:2018} and we predicted here similar values for \ce{PaF^3+}.  
Due to the large differences in energy of the levels in Yb, the absolute values
of the enhancement factors $K$ are reduced to below 15. In molecular systems, however, 
orders of magnitude higher absolute values of $K$ are in principle found
regularly as a result of the quasi degenerate vibronic states.

Measuring corresponding transition wavenumber very accurately is in principle
possible under favourable circumstances with the help of frequency combs
(see e.g.\ Refs.~\cite{udem:2002} and \cite{flambaum:2007}).

This requires, however, narrow linewidths in
order to detect corresponding frequency shifts of these lines due
to a variation of fundamental constants. Frequency shifts are considered to be resolvable to about $10^{-3}\times
\Gamma$ to $10^{-6}\times \Gamma$ \cite{beloy:2011b}. For our current estimates of the
linewidths we restricted ourselves to natural radiative linewidths
of the states involved as obtained within the electric dipole approximation (see
computational details). For a more accurate estimate, however, higher order
terms (magnetic dipole, electric quadrupole, etc.) 
as well as additional broadening effects on the absorption
linewidths would have to be included.

\subsubsection{\ce{CeF^2+}}

From the vibrationally corrected  $\Delta q_{1,i}$ values ($\Delta q_{1,iv}$) and
the electronic and vibrational transition wavenumbers we computed the enhancement
factors $K_{\alpha,iv,jw}$, $K_{\mu,iv,iw}$, the
natural radiative line widths $\Gamma_{iv}, \Gamma_{jw}$ for
vibronic states $iv, jw$ and the expected frequency shifts. We
ranked the results with decreasing value for $\left| K_{\alpha}\right|$ and give
the corresponding level pairs in
Tab.~\ref{tab:vibstates_ksort_ce}. From this we find very large enhancement
factors $\left| K_{\alpha}\right| \gg1000$ for about ten level pairs. For
comparison, the enhancement of the most favourable transition in \ce{Yb+}, which
provides current best limits on $\delta\alpha$, is $K=-5.95$
\cite{safronova:2018}. Large absolute values for the enhancement
were found previously in high-lying vibronic states of molecular
ions such as \ce{I2+} \cite{pasteka:2015} but with smaller individual $\left|
\Delta q_1 \right|$ values.

We see large $\left| \Delta q_1 \right|$ values for transitions into the higher
lying $(3)3/2$ state and, thus, also for $\left| K \right|$ as expected from
Tab.~\ref{tab:cghf_dhf} and \ref{tab:paf}.
These transitions, however, involve states with
quite large
radiative linewidths. 
The radiative linewidth of lower lying electronic states
are expected to be much smaller at the cost of a reduced
enhancement $\left| K \right|$.

The present Hartree--Fock approach is not highly accurate, and we further want to
note that in this ansatz we only include anharmonicities in the vibrational levels
within the DVR approach and consider no geometry dependence due to the variation
of $\alpha$. Thus, we can only hint in this work to the numerous possibilities of
close lying levels with very large enhancement factors $\left|K_{\alpha}\right|$
and coarsely estimate the corresponding linewidths. But for which transitions
precisely those near degeneracies emerge depends sensitively on the detailed
alignment of electronic states as well as the shape of the potential energy
profiles that determine the vibrational level structure, which requires the use of
more demanding higher-level computational methods to be predicted reliably.

Corresponding to our presently chosen model, we provide expected wavenumber shifts
employing current best limits on temporal variations of $\delta\mu/\mu$ and
$\delta\alpha/\alpha$ along with estimated linewidths for pairs of vibronic
transitions in Tab.~\ref{tab:vibstates_ksort_ce}. Expected frequency
shifts are obtained as (this is exemplified for the transition
$(1)7/2(40)\leftarrow
(\mathrm{X})5/2(44)$) 

\begin{multline*}
\delta \tilde{\nu} < \SI{0.26}{\per\centi\meter} \\
 \times \parantheses{16000\times 10^{-18}\,\si{\per
yr}-\SI{4500e-17}{\per yr}}\\
< -7.15\times 10^{-15}~\si{\mathrm{cm}^{-1}\per yr}=-0.2\,\mathrm{mHz}\, c^{-1}/\mathrm{yr}\,,
\end{multline*}
where we assume the current best limits for temporal variations of $\alpha$ and
$\mu$: $\delta\mu/\mu<10^{-17}\,\si{\per yr}$ and
$\delta\alpha/\alpha<10^{-18}\,\si{\per yr}$.  The relative wavenumber shift of
$\delta\tilde{\nu}$ on the transition wavenumber is on the order of
$10^{-13}$ and appears in principle measurable with modern
high-precision molecular spectroscopy \cite{demille:2008}. However, even much
larger absolute enhancements may be possible depending on the exact energetic
position of the vibronic states which is currently only crudely described on the
HF level.  Therefore, we anticipate that high-precision spectroscopy has the
potential to improve current bounds on VFC in the near future. We summarize these
results in Fig.~\ref{fig:fancy_ce} and conclude that \ce{CeF^2+} can be considered
a promising candidate for future precision experiments that aim to detect VFC. 

\subsubsection{\ce{PaF^3+}}

\begin{table}
\begin{threeparttable}
\caption{
Analytically determined results for the
sensitivity parameters $\Delta q_{1}$ and $2\Delta q_{2}$ for \ce{PaF^3+} computed on the level of
ZORA-cGHF. 
$\Delta q_i$ are computed with the electronic ground state as reference as $\Delta q_i = q_i\left(
\mathrm{excited\ state} \right) - q_i\left( \mathrm{ground\ state} \right)$.
For each electronic state the corresponding equilibrium bond length was used
as internuclear distance (see Supplemental Material).
}
\label{tab:paf}
\begin{tabular}{l 
S[table-format=-5.3,round-mode=places,round-precision=1]
S[table-format=-5.3,round-mode=places,round-precision=0]
}
\toprule
{State} & 
{$\Delta q_{1}/\si{cm^{-1}}$ } &   
{$2\Delta q_{2}/\si{cm^{-1}}$ } \\  
\midrule
(1)3/2  & 738.29    & -1426.13 \\
(1)1/2  & -2101.59  & -5365.66 \\
(1)7/2  & 3473.55   & -4909.71 \\
(1)5/2  & 4341.24   & -5645.34 \\
(2)3/2  & 2806.47   & -9261.39 \\
(2)1/2  & 690.11    & -9578.94 \\
(3)3/2  & -42839    & -1570.90 \\
\bottomrule
\end{tabular}
\end{threeparttable}
\end{table}

The enhancement $\Delta q_{1}$ and $\Delta q_{2}$ of $\alpha$ variations  in
\ce{PaF^3+} is given in Tab.~\ref{tab:paf}. As for \ce{CeF^2+} we show
the 40 vibronic transitions with largest absolute enhancement factors $\left|
K_{\alpha}\right|$ alongside the linewidth and expected frequency shifts
in Tab.~\ref{tab:vibstates_ksort_pa}). 
From the data in Tab.~\ref{tab:vibstates_ksort_pa} a wide variety of states with different transition
wavenumbers and with large absolute sensitivity parameters is observed.
An exemplary estimation of the
expected frequency shift for the transition  $(1)7/2(11),(\mathrm{X})5/2(18)$
is given as

\begin{multline*}
\delta \tilde{\nu} < \SI{0.49}{\per\centi\meter} \\
 \times \parantheses{13000\times 10^{-18}\,\si{\per
yr}-\SI{5600e-17}{\per
yr}}\\
< -2.1\times 10^{-14}~\si{cm^{-1}\per yr}=-0.6\,\mathrm{mHz}\, c^{-1}/\mathrm{yr}
\,,
\end{multline*}
with the same assumptions and limitations to our computational approach as in the case of \ce{CeF^2+}.

\ce{PaF^3+} generally exhibits transitions with larger values for $\left| \Delta
q_{1}\right|$ than \ce{CeF^2+}. The transitions with highest absolute sensitivity
involve the $(3)3/2$ state, which is expected from the discussion above.

Computed natural radiative linewidths in \ce{PaF^3+} are similar to those
in \ce{CeF^2+} but
slightly broader, which aligns with the larger mixing of states with
different orbital angular momenta via spin-orbit coupling. Nevertheless, the same
arguments (and limitations regarding accuracy of the estimates) as for \ce{CeF^2+} still apply.

Although most of the $\left| \Delta q_1\right|$ values in \ce{PaF^3+} are larger in magnitude than in
\ce{CeF^2+},
we observe different transition with predicted large absolute shifts
$\delta\tilde{\nu}$ in both compounds. This highlights the various
possibilities to obtain limits on variations of $\alpha$ and $\mu$ at
the same time.  

\section{Conclusion}
The tedious numerical procedure needed previously to assess the
impact of a potential variation of fundamental constants likely hampered
routine investigations. 
We introduced herein an efficient way to compute the sensitivity of
atoms and molecules to a variation of the fine-structure constant $\alpha$ via an
expectation value of operators that are available in essentially all common 
\textit{ab initio} relativistic
electronic structure programs. By comparison to the conventional numerical
approach we demonstrated the advantages of our analytic scheme
of computation, which allows to routinely estimate sensitivities to a variation of
$\alpha$.

As the corresponding expectation values can possibly be found reported in legacy outputs
of previous calculations, this approach has the potential to open up access to a rich source of
valuable data.

We applied our scheme herein to find huge enhancements of
variations of $\alpha$ in the highly charged molecules
\ce{PaF^3+} and \ce{CeF^2+}. Both systems are expected to be
well-controllable and possess energetically close-lying electronic states of different
symmetry, by which many accidentally degenerate vibronic
levels emerge that display also large enhancements of changes in $\alpha$.
This highlights that stable polar highly charged
molecular ions such as \ce{PaF^3+} or \ce{CeF^2+} are well posed
for searches for a temporal variation of fundamental constants.

\begin{acknowledgments}
We thank J\"urgen Stohner and Stephan Malbrunot-Ettenauer for discussions.
For the computational work, we acknowledge computing time provided 
at the NHR Center NHR@SW at
Goethe-University Frankfurt. This was funded by the
Federal Ministry of Education and Research and the state
governments participating on the basis of the resolutions 
of the GWK for national high performance computing at universities 
(http://www.nhr-verein.de/unsere-partner).
C.Z. and R.B. acknowledge funding received from the Deutsche
Forschungsgemeinschaft (DFG, German Research Foundation) -- Projektnummer
445296313. 
\end{acknowledgments}

\clearpage

\begin{table*}
\resizebox{\textwidth}{!}{%
\begin{threeparttable}
\caption{Accidentally nearly degenerate vibronic states
with transition wavenumber $\tilde{\nu}$
in \ce{CeF^2+} computed on the level of ZORA-cGHF are shown. The
results are ranked according to the enhancement factor $
K_{\alpha}$. Corresponding electronic sensitivity factors $\Delta q_1$, enhancement
factors $K_{\mu}$ for the variation of the electron-proton
mass-ratio, the natural radiative linewidths $\Gamma$ of the pair of states
and expected frequency shifts $\delta\nu$ including
the individual contributions from $\delta \alpha$ and $\delta\mu$ are
provided.  Expected frequency shifts were computed assuming the
currently best limits on $\delta \alpha$ and
$\delta\mu$\cite{godun:2014,huntemann:2014,lange:2021}:
$\delta\mu/\mu<10^{-17}\,\si{\per yr}$ and
$\delta\alpha/\alpha<10^{-18}\,\si{\per yr}$. The values in
parentheses after the term symbol indicate the vibrational quantum
number.}
\label{tab:vibstates_ksort_ce}
\begin{tabular}{
l 
c
l
S[table-format=-2.2,round-mode=figures,round-precision=2]
S[table-format=-1.1e-2,round-mode=figures,round-precision=2,scientific-notation = true]
S[table-format=-1.1e-2,round-mode=figures,round-precision=2,scientific-notation = true]
S[table-format=-1.1e-2,round-mode=figures,round-precision=2,scientific-notation = true]
S[table-format=-2.0e-2,round-mode=figures,round-precision=1]
S[table-format=-2.0e-2,round-mode=figures,round-precision=1]
S[table-format=-1.2,round-mode=places,round-precision=2]
S[table-format=-1.2,round-mode=places,round-precision=2]
S[table-format=-1.2,round-mode=figures,round-precision=2]
}
\toprule
\multicolumn{3}{c}{Transition}                                      &  
{\multirow{2}{*}{$\tilde{\nu} / \si{cm^{-1}}$}           }  &    
{\multirow{2}{*}{$\Delta q_{1} / \si{cm^{-1}}$}                       }  & 
{\multirow{2}{*}{$ K_{\alpha}$}                                }  &
{\multirow{2}{*}{$ K_{\mu}$}                                   }  &
{\multirow{2}{*}{$\Gamma^\mathrm{A} / \si{Hz}$}             }  &
{\multirow{2}{*}{$\Gamma^\mathrm{B} / \si{Hz}$}             }  &
{\multirow{2}{*}{$\delta \nu / \si{mHz}$}        }  &
{\multirow{2}{*}{$\delta \nu_{\alpha} / \si{mHz}$} }  &
{\multirow{2}{*}{$\delta \nu_{\mu} / \si{mHz}$}    }  \\
\multicolumn{1}{c}{A} &  & \multicolumn{1}{c}{B} \\
\midrule
(1)7/2(67)  &         $\leftarrow$  &         (3)3/2(16)  &         2.528   &       25926.573   &       20515.193   &     5740.971   &  4.95806e+02  &  1.36265e+07  &  5.90468   &  1.55452   &  4.35016   \\
(3)3/2(11)  &         $\leftarrow$  &         (1)7/2(60)  &         2.662   &       -25940.206  &       -19491.135  &     -5450.557  &  1.31653e+07  &  4.67473e+02  &  -5.90472  &  -1.55534  &  -4.34939  \\
(3)3/2(0)   &         $\leftarrow$  &         (1)7/2(45)  &         2.853   &       -25969.321  &       -18204.167  &     -5084.925  &  1.21256e+07  &  3.95578e+02  &  -5.90644  &  -1.55708  &  -4.34936  \\
(1)3/2(60)  &         $\leftarrow$  &         (3)3/2(9)   &         3.017   &       23996.181   &       15907.808   &     5018.603   &  4.53055e+02  &  1.29754e+07  &  5.97783   &  1.43877   &  4.53905   \\
(1)7/2(40)  &         $\leftarrow$  &         (X)5/2(44)  &         0.257   &       2034.217    &       15812.479   &     -4477.836  &  3.66614e+02  &  3.86760e+02  &  -0.22343  &  0.12197   &  -0.34540  \\
(1)1/2(37)  &         $\leftarrow$  &         (2)3/2(33)  &         0.526   &       -2264.992   &       -8620.162   &     2192.802   &  3.57060e+02  &  3.58379e+02  &  0.20966   &  -0.13581  &  0.34546   \\
(3)3/2(1)   &         $\leftarrow$  &         (2)1/2(44)  &         11.882  &       -25666.526  &       -4320.076   &     -1152.286  &  1.22126e+07  &  4.33941e+02  &  -5.64368  &  -1.53893  &  -4.10475  \\
(3)3/2(17)  &         $\leftarrow$  &         (2)3/2(66)  &         13.432  &       -25902.885  &       -3856.932   &     -1025.368  &  1.37166e+07  &  5.53117e+02  &  -5.68202  &  -1.55310  &  -4.12892  \\
(3)3/2(4)   &         $\leftarrow$  &         (1)3/2(53)  &         12.739  &       -24017.162  &       -3770.613   &     -1187.897  &  1.24965e+07  &  4.24273e+02  &  -5.97672  &  -1.44003  &  -4.53669  \\
(1)1/2(61)  &         $\leftarrow$  &         (1)5/2(57)  &         1.324   &       -2348.230   &       -3547.918   &     695.069    &  4.84654e+02  &  1.02350e+03  &  0.13504   &  -0.14080  &  0.27583   \\
(3)3/2(17)  &         $\leftarrow$  &         (2)1/2(66)  &         15.528  &       -25633.305  &       -3301.569   &     -881.646   &  1.37166e+07  &  5.66670e+02  &  -5.64114  &  -1.53693  &  -4.10420  \\
(X)5/2(60)  &         $\leftarrow$  &         (1)5/2(54)  &         1.404   &       -2134.116   &       -3039.017   &     1175.016   &  4.62850e+02  &  1.01161e+03  &  0.36678   &  -0.12796  &  0.49474   \\
(3)3/2(12)  &         $\leftarrow$  &         (2)1/2(59)  &         17.279  &       -25641.421  &       -2968.015   &     -792.272   &  1.32594e+07  &  5.21245e+02  &  -5.64136  &  -1.53742  &  -4.10394  \\
(3)3/2(5)   &         $\leftarrow$  &         (X)5/2(56)  &         16.661  &       -23918.722  &       -2871.286   &     -939.534   &  1.25936e+07  &  4.47384e+02  &  -6.12685  &  -1.43413  &  -4.69272  \\
(3)3/2(0)   &         $\leftarrow$  &         (1)1/2(47)  &         16.863  &       -23717.959  &       -2813.001   &     -884.948   &  1.21256e+07  &  4.18062e+02  &  -5.89588  &  -1.42209  &  -4.47379  \\
(1)5/2(57)  &         $\leftarrow$  &         (3)3/2(10)  &         19.590  &       26036.328   &       2658.142    &     715.766    &  1.02350e+03  &  1.30706e+07  &  5.76472   &  1.56110   &  4.20362   \\
(1)1/2(61)  &         $\leftarrow$  &         (3)3/2(10)  &         20.914  &       23688.098   &       2265.330    &     714.456    &  4.84654e+02  &  1.30706e+07  &  5.89976   &  1.42030   &  4.47945   \\
(2)3/2(55)  &         $\leftarrow$  &         (3)3/2(9)   &         24.953  &       25929.303   &       2078.214    &     552.701    &  4.87481e+02  &  1.29754e+07  &  5.68936   &  1.55468   &  4.13468   \\
(1)7/2(44)  &         $\leftarrow$  &         (1)1/2(46)  &         2.600   &       2251.386    &       1732.165    &     -161.868   &  3.89930e+02  &  4.12377e+02  &  0.00884   &  0.13499   &  -0.12615  \\
(1)3/2(17)  &         $\leftarrow$  &         (1)7/2(15)  &         2.267   &       -1906.892   &       -1682.395   &     278.339    &  1.82493e+02  &  1.67375e+02  &  0.07482   &  -0.11433  &  0.18916   \\
(3)3/2(3)   &         $\leftarrow$  &         (1)7/2(49)  &         31.412  &       -25958.343  &       -1652.765   &     -461.403   &  1.24006e+07  &  4.17550e+02  &  -5.90150  &  -1.55642  &  -4.34508  \\
(3)3/2(18)  &         $\leftarrow$  &         (1)5/2(68)  &         34.011  &       -26000.155  &       -1528.918   &     -411.481   &  1.38063e+07  &  1.06499e+03  &  -5.75452  &  -1.55893  &  -4.19559  \\
(1)1/2(54)  &         $\leftarrow$  &         (3)3/2(5)   &         31.197  &       23699.113   &       1519.340    &     479.123    &  4.54794e+02  &  1.25936e+07  &  5.90196   &  1.42096   &  4.48100   \\
(3)3/2(10)  &         $\leftarrow$  &         (X)5/2(63)  &         32.293  &       -23904.584  &       -1480.479   &     -484.483   &  1.30706e+07  &  4.74897e+02  &  -6.12366  &  -1.43328  &  -4.69038  \\
(3)3/2(12)  &         $\leftarrow$  &         (2)3/2(59)  &         39.651  &       -25921.422  &       -1307.480   &     -347.015   &  1.32594e+07  &  5.07299e+02  &  -5.67920  &  -1.55421  &  -4.12499  \\
(3)3/2(14)  &         $\leftarrow$  &         (1)3/2(67)  &         39.203  &       -23975.741  &       -1223.146   &     -385.669   &  1.34445e+07  &  4.80496e+02  &  -5.97027  &  -1.43755  &  -4.53273  \\
(3)3/2(4)   &         $\leftarrow$  &         (2)1/2(48)  &         43.587  &       -25653.284  &       -1177.100   &     -313.764   &  1.24965e+07  &  4.58374e+02  &  -5.63813  &  -1.53813  &  -4.10000  \\
(3)3/2(13)  &         $\leftarrow$  &         (1)5/2(61)  &         49.358  &       -26023.993  &       -1054.491   &     -283.382   &  1.33523e+07  &  1.03931e+03  &  -5.75365  &  -1.56036  &  -4.19329  \\
(X)5/2(66)  &         $\leftarrow$  &         (3)3/2(12)  &         46.871  &       23899.517   &       1019.796    &     334.641    &  4.86558e+02  &  1.32594e+07  &  6.13522   &  1.43298   &  4.70224   \\
(3)3/2(15)  &         $\leftarrow$  &         (1)1/2(68)  &         47.573  &       -23677.510  &       -995.424    &     -313.365   &  1.35359e+07  &  5.12743e+02  &  -5.88886  &  -1.41967  &  -4.46919  \\
(3)3/2(2)   &         $\leftarrow$  &         (1)5/2(46)  &         55.387  &       -26068.148  &       -941.301    &     -252.481   &  1.23041e+07  &  9.72980e+02  &  -5.75539  &  -1.56301  &  -4.19238  \\
(2)1/2(55)  &         $\leftarrow$  &         (3)3/2(9)   &         56.149  &       25645.705   &       913.487     &     244.456    &  5.01136e+02  &  1.29754e+07  &  5.65263   &  1.53768   &  4.11495   \\
(2)1/2(51)  &         $\leftarrow$  &         (1)3/2(56)  &         3.686   &       1641.115    &       890.569     &     -390.549   &  4.77068e+02  &  4.37447e+02  &  -0.33312  &  0.09840   &  -0.43152  \\
(2)1/2(25)  &         $\leftarrow$  &         (X)5/2(31)  &         3.905   &       1733.362    &       887.811     &     -502.380   &  2.90715e+02  &  3.04274e+02  &  -0.48417  &  0.10393   &  -0.58810  \\
(3)3/2(1)   &         $\leftarrow$  &         (2)3/2(44)  &         58.550  &       -25964.825  &       -886.921    &     -234.841   &  1.22126e+07  &  4.27416e+02  &  -5.67897  &  -1.55681  &  -4.12216  \\
(1)3/2(12)  &         $\leftarrow$  &         (2)3/2(8)   &         4.868   &       -1935.079   &       -795.081    &     279.861    &  1.33584e+02  &  1.31683e+02  &  0.29237   &  -0.11602  &  0.40840   \\
(X)5/2(22)  &         $\leftarrow$  &         (1)5/2(17)  &         5.572   &       -2172.864   &       -779.861    &     296.525    &  2.30465e+02  &  7.56047e+02  &  0.36509   &  -0.13028  &  0.49537   \\
(1)5/2(64)  &         $\leftarrow$  &         (3)3/2(15)  &         67.299  &       26013.922   &       773.085     &     208.705    &  1.05148e+03  &  1.35359e+07  &  5.77053   &  1.55976   &  4.21077   \\
(1)3/2(49)  &         $\leftarrow$  &         (3)3/2(1)   &         62.944  &       24033.522   &       763.643     &     241.016    &  4.04706e+02  &  1.22126e+07  &  5.98905   &  1.44101   &  4.54804   \\
(3)3/2(0)   &         $\leftarrow$  &         (X)5/2(49)  &         64.345  &       -23937.577  &       -744.043    &     -242.901   &  1.21256e+07  &  4.13936e+02  &  -6.12083  &  -1.43526  &  -4.68557  \\
(1)7/2(56)  &         $\leftarrow$  &         (3)3/2(8)   &         70.414  &       25947.775   &       737.009     &     206.558    &  4.51663e+02  &  1.28796e+07  &  5.91613   &  1.55579   &  4.36034   \\
(2)1/2(69)  &         $\leftarrow$  &         (3)3/2(19)  &         69.614  &       25630.429   &       736.361     &     197.270    &  5.84620e+02  &  1.38960e+07  &  5.65373   &  1.53676   &  4.11697   \\
(1)1/2(31)  &         $\leftarrow$  &         (2)1/2(27)  &         5.330   &       -1952.209   &       -732.522    &     231.917    &  3.14472e+02  &  3.08839e+02  &  0.25354   &  -0.11705  &  0.37059   \\
(1)3/2(58)  &         $\leftarrow$  &         (2)3/2(53)  &         5.340   &       -1930.983   &       -723.182    &     255.138    &  4.45345e+02  &  4.76617e+02  &  0.29269   &  -0.11578  &  0.40847   \\
(1)5/2(53)  &         $\leftarrow$  &         (3)3/2(7)   &         72.592  &       26046.593   &       717.621     &     193.525    &  1.00745e+03  &  1.27842e+07  &  5.77328   &  1.56171   &  4.21157   \\
(3)3/2(6)   &         $\leftarrow$  &         (1)7/2(53)  &         74.720  &       -25952.152  &       -694.647    &     -193.681   &  1.26883e+07  &  4.38183e+02  &  -5.89464  &  -1.55605  &  -4.33858  \\
(2)3/2(51)  &         $\leftarrow$  &         (3)3/2(6)   &         76.466  &       25938.236   &       678.425     &     180.702    &  4.64878e+02  &  1.26883e+07  &  5.69762   &  1.55522   &  4.14240   \\
(2)3/2(31)  &         $\leftarrow$  &         (X)5/2(37)  &         6.070   &       2049.697    &       675.349     &     -309.588   &  3.44028e+02  &  3.45245e+02  &  -0.44048  &  0.12290   &  -0.56337  \\
(1)1/2(22)  &         $\leftarrow$  &         (1)5/2(19)  &         7.262   &       -2390.600   &       -658.368    &     127.103    &  2.38022e+02  &  7.74922e+02  &  0.13339   &  -0.14334  &  0.27672   \\
(3)3/2(8)   &         $\leftarrow$  &         (1)1/2(58)  &         73.264  &       -23692.644  &       -646.774    &     -203.302   &  1.28796e+07  &  4.72121e+02  &  -5.88591  &  -1.42058  &  -4.46534  \\
(2)3/2(69)  &         $\leftarrow$  &         (3)3/2(19)  &         80.128  &       25894.698   &       646.334     &     172.467    &  5.69732e+02  &  1.38960e+07  &  5.69555   &  1.55261   &  4.14295   \\
(1)5/2(20)  &         $\leftarrow$  &         (1)1/2(23)  &         7.447   &       2389.466    &       641.726     &     -122.961   &  7.84116e+02  &  2.47088e+02  &  -0.13125  &  0.14327   &  -0.27452  \\
(X)5/2(52)  &         $\leftarrow$  &         (3)3/2(2)   &         75.021  &       23928.708   &       637.922     &     209.263    &  4.29310e+02  &  1.23041e+07  &  6.14119   &  1.43473   &  4.70646   \\
(3)3/2(4)   &         $\leftarrow$  &         (2)3/2(48)  &         82.191  &       -25945.816  &       -631.358    &     -167.151   &  1.24965e+07  &  4.48787e+02  &  -5.67429  &  -1.55567  &  -4.11862  \\
(2)3/2(62)  &         $\leftarrow$  &         (3)3/2(14)  &         82.452  &       25914.105   &       628.589     &     167.620    &  5.27444e+02  &  1.34445e+07  &  5.69707   &  1.55377   &  4.14329   \\
(3)3/2(5)   &         $\leftarrow$  &         (1)5/2(50)  &         85.058  &       -26054.239  &       -612.624    &     -164.235   &  1.25936e+07  &  9.93919e+02  &  -5.75011  &  -1.56217  &  -4.18794  \\
(1)3/2(43)  &         $\leftarrow$  &         (1)5/2(39)  &         6.706   &       -2048.962   &       -611.112    &     168.591    &  3.73050e+02  &  9.31099e+02  &  0.21607   &  -0.12285  &  0.33892   \\ 
\bottomrule
\end{tabular}
\end{threeparttable}
}
\end{table*}

%%%%%%%%%%%%%%%%%%%%%%%%%%%%%%%%%%%%%%%%%%%%%%%%%%%%%%%%%%%%%%%%%%%%%%%%%%%%%%%%%%%%%%%%%%%%%%%%%%%%%%%%%%%%%%%%%%%%%%%%

\begin{table*}
\resizebox{\textwidth}{!}{%
\begin{threeparttable}
\caption{Accidentally nearly degenerate vibronic transition pairs
with transition wavenumber difference $\tilde{\nu}$
in \ce{PaF^3+} computed on the level of ZORA-cGHF
are shown. The results are ranked according to the enhancement
factor $K_{\alpha}$. Corresponding electronic sensitivity
factors $\Delta q_1$, enhancement
factors $ K_{\mu}$ for the variation of the electron-proton
mass-ratio,
the natural radiative linewidths $\Gamma$ of the pair of transitions and
expected frequency shifts $\delta\nu$ including
the individual contributions from $\delta \alpha$ and $\delta\mu$ are provided.
Expected frequency shifts were computed assuming the currently best
limits on $\delta \alpha$ and $\delta\mu$\cite{godun:2014,huntemann:2014,lange:2021}:
$\delta\mu/\mu<10^{-17}\,\si{\per yr}$ and
$\delta\alpha/\alpha<10^{-18}\,\si{\per yr}$. The values in parentheses after the
term symbol indicate the vibrational quantum number.}
\label{tab:vibstates_ksort_pa}
\begin{tabular}{
l 
c
l
S[table-format=-2.2e-2,round-mode=figures,round-precision=2]
S[table-format=-1.1e-2,round-mode=figures,round-precision=2,scientific-notation = true]
S[table-format=-1.1e-2,round-mode=figures,round-precision=2,scientific-notation = true]
S[table-format=-1.1e-2,round-mode=figures,round-precision=2,scientific-notation = true]
S[table-format=-2.0e-2,round-mode=figures,round-precision=1]
S[table-format=-2.0e-2,round-mode=figures,round-precision=1]
S[table-format=-1.2,round-mode=places,round-precision=2]
S[table-format=-1.2,round-mode=places,round-precision=2]
S[table-format=-1.2,round-mode=figures,round-precision=2]
}
\toprule
\multicolumn{3}{c}{Transition}                                       & 
{\multirow{2}{*}{$\tilde{\nu} / \si{cm^{-1}}$}           }  &    
{\multirow{2}{*}{$\Delta q_{1} / \si{cm^{-1}}$}                       }  & 
{\multirow{2}{*}{$ K_{\alpha}$}                                }  &
{\multirow{2}{*}{$ K_{\mu}$}                                   }  &
{\multirow{2}{*}{$\Gamma^\mathrm{A} / \si{Hz}$}             }  &
{\multirow{2}{*}{$\Gamma^\mathrm{B} / \si{Hz}$}             }  &
{\multirow{2}{*}{$\delta \nu / \si{mHz}$}        }  &
{\multirow{2}{*}{$\delta \nu_{\alpha} / \si{mHz}$} }  &
{\multirow{2}{*}{$\delta \nu_{\mu} / \si{mHz}$}    }  \\
\multicolumn{1}{c}{A} &  & \multicolumn{1}{c}{B} \\
\midrule
(2)1/2(6)   &         $\leftarrow$  &         (1)7/2(10)  &         0.059    &       -2765.299   &       -94167.341  &     -28092.336  &  3.53414e+03  &  1.75178e+02  &  -0.66043  &  -0.16580  &  -0.49463  \\
(1)7/2(11)  &         $\leftarrow$  &         (X)5/2(18)  &         0.492    &       3322.453    &       13493.832   &     -5608.213   &  1.85585e+02  &  1.91805e+02  &  -0.62873  &  0.19921   &  -0.82794  \\
(1)3/2(29)  &         $\leftarrow$  &         (1)5/2(22)  &         0.565    &       -3577.435   &       -12672.123  &     4592.151    &  2.68906e+02  &  1.30311e+05  &  0.56280   &  -0.21450  &  0.77730   \\
(2)1/2(24)  &         $\leftarrow$  &         (1)5/2(27)  &         0.624    &       -3405.407   &       -10908.411  &     -1919.170   &  3.78673e+03  &  1.32508e+05  &  -0.56341  &  -0.20418  &  -0.35923  \\
(2)1/2(8)   &         $\leftarrow$  &         (2)3/2(9)   &         0.538    &       -2076.395   &       -7712.033   &     -764.764    &  3.56521e+03  &  9.99198e+04  &  -0.24796  &  -0.12450  &  -0.12346  \\
(X)5/2(41)  &         $\leftarrow$  &         (3)3/2(2)   &         18.807   &       42574.857   &       4527.633    &     790.456     &  3.61576e+02  &  1.24448e+07  &  7.00939   &  2.55272   &  4.45667   \\
(1)1/2(34)  &         $\leftarrow$  &         (2)1/2(26)  &         1.525    &       -2903.077   &       -3807.671   &     1902.480    &  5.60169e+02  &  3.81179e+03  &  0.69564   &  -0.17406  &  0.86970   \\
(3)3/2(3)   &         $\leftarrow$  &         (2)1/2(30)  &         24.796   &       -43292.753  &       -3491.902   &     -420.717    &  1.25210e+07  &  3.85983e+03  &  -5.72325  &  -2.59577  &  -3.12748  \\
(3)3/2(6)   &         $\leftarrow$  &         (X)5/2(46)  &         25.071   &       -42485.849  &       -3389.226   &     -592.072    &  1.27539e+07  &  3.93131e+02  &  -6.99748  &  -2.54739  &  -4.45009  \\
(1)5/2(26)  &         $\leftarrow$  &         (2)1/2(23)  &         2.087    &       3424.319    &       3280.874    &     574.682     &  1.32051e+05  &  3.77398e+03  &  0.56495   &  0.20532   &  0.35964   \\
(2)3/2(8)   &         $\leftarrow$  &         (2)1/2(7)   &         1.322    &       2085.904    &       3156.486    &     312.290     &  9.98994e+04  &  3.54980e+03  &  0.24880   &  0.12507   &  0.12374   \\
(3)3/2(3)   &         $\leftarrow$  &         (1)5/2(33)  &         38.342   &       -46587.845  &       -2430.091   &     -303.161    &  1.25210e+07  &  1.35316e+05  &  -6.27811  &  -2.79334  &  -3.48477  \\
(2)1/2(25)  &         $\leftarrow$  &         (1)5/2(28)  &         3.031    &       -3386.869   &       -2234.703   &     -394.917    &  3.79934e+03  &  1.32972e+05  &  -0.56194  &  -0.20307  &  -0.35887  \\
(1)1/2(34)  &         $\leftarrow$  &         (1)5/2(29)  &         6.687    &       -6271.769   &       -1875.802   &     254.974     &  5.60169e+02  &  1.33440e+05  &  0.13511   &  -0.37605  &  0.51115   \\
(2)1/2(9)   &         $\leftarrow$  &         (2)3/2(10)  &         2.403    &       -2066.678   &       -1719.880   &     -170.966    &  3.58043e+03  &  9.99390e+04  &  -0.24709  &  -0.12391  &  -0.12318  \\
(3)3/2(3)   &         $\leftarrow$  &         (2)3/2(31)  &         66.684   &       -45142.330  &       -1353.919   &     -162.307    &  1.25210e+07  &  9.99292e+04  &  -5.95140  &  -2.70667  &  -3.24474  \\
(1)5/2(25)  &         $\leftarrow$  &         (2)1/2(22)  &         5.114    &       3443.509    &       1346.731    &     234.876     &  1.31599e+05  &  3.76107e+03  &  0.56656   &  0.20647   &  0.36009   \\
(1)7/2(32)  &         $\leftarrow$  &         (3)3/2(1)   &         68.438   &       45811.621   &       1338.779    &     177.222     &  3.67387e+02  &  1.23672e+07  &  6.38289   &  2.74680   &  3.63609   \\
(2)1/2(26)  &         $\leftarrow$  &         (1)5/2(29)  &         5.162    &       -3368.692   &       -1305.146   &     -231.684    &  3.81179e+03  &  1.33440e+05  &  -0.56053  &  -0.20198  &  -0.35855  \\
(2)3/2(7)   &         $\leftarrow$  &         (2)1/2(6)   &         3.218    &       2095.221    &       1302.094    &     128.546     &  9.98778e+04  &  3.53414e+03  &  0.24965   &  0.12563   &  0.12402   \\
(2)3/2(30)  &         $\leftarrow$  &         (3)3/2(2)   &         76.892   &       45156.031   &       1174.537    &     141.694     &  9.99609e+04  &  1.24448e+07  &  5.97375   &  2.70749   &  3.26626   \\
(3)3/2(2)   &         $\leftarrow$  &         (1)7/2(33)  &         84.153   &       -45793.192  &       -1088.334   &     -143.220    &  1.24448e+07  &  3.73917e+02  &  -6.35891  &  -2.74569  &  -3.61322  \\
(1)7/2(0)   &         $\leftarrow$  &         (1)1/2(3)   &         10.666   &       5480.771    &       1027.722    &     -116.726    &  6.11653e+01  &  2.12617e+02  &  -0.04462  &  0.32862   &  -0.37324  \\
(1)5/2(7)   &         $\leftarrow$  &         (X)5/2(15)  &         8.584    &       4315.346    &       1005.440    &     -373.838    &  1.26630e+05  &  1.63829e+02  &  -0.70330  &  0.25874   &  -0.96204  \\
(2)1/2(10)  &         $\leftarrow$  &         (2)3/2(11)  &         4.253    &       -2056.792   &       -967.273    &     -96.397     &  3.59539e+03  &  9.99555e+04  &  -0.24622  &  -0.12332  &  -0.12290  \\
(2)1/2(27)  &         $\leftarrow$  &         (1)5/2(30)  &         7.095    &       -3350.667   &       -944.533    &     -168.435    &  3.82405e+03  &  1.33910e+05  &  -0.55916  &  -0.20090  &  -0.35826  \\
(1)5/2(32)  &         $\leftarrow$  &         (3)3/2(2)   &         105.573  &       46610.348   &       882.995     &     110.785     &  1.34853e+05  &  1.24448e+07  &  6.30103   &  2.79469   &  3.50634   \\
(1)5/2(24)  &         $\leftarrow$  &         (2)1/2(21)  &         8.453    &       3462.945    &       819.361     &     142.296     &  1.31157e+05  &  3.74803e+03  &  0.56822   &  0.20763   &  0.36059   \\
(2)3/2(6)   &         $\leftarrow$  &         (2)1/2(5)   &         5.140    &       2104.326    &       818.883     &     80.679      &  9.98540e+04  &  3.51829e+03  &  0.25048   &  0.12617   &  0.12431   \\
(2)1/2(29)  &         $\leftarrow$  &         (3)3/2(2)   &         116.622  &       43296.441   &       742.511     &     90.059      &  3.84803e+03  &  1.24448e+07  &  5.74467   &  2.59599   &  3.14868   \\
(2)1/2(28)  &         $\leftarrow$  &         (1)5/2(31)  &         8.982    &       -3332.478   &       -742.058    &     -132.946    &  3.83614e+03  &  1.34382e+05  &  -0.55779  &  -0.19981  &  -0.35798  \\
(2)1/2(11)  &         $\leftarrow$  &         (2)3/2(12)  &         6.072    &       -2046.738   &       -674.154    &     -67.365     &  3.61017e+03  &  9.99712e+04  &  -0.24535  &  -0.12272  &  -0.12263  \\
(1)7/2(9)   &         $\leftarrow$  &         (2)1/2(5)   &         8.477    &       2774.607    &       654.658     &     195.148     &  1.64583e+02  &  3.51829e+03  &  0.66227   &  0.16636   &  0.49591   \\
(1)3/2(30)  &         $\leftarrow$  &         (2)3/2(21)  &         6.260    &       -2040.730   &       -652.026    &     540.306     &  2.75822e+02  &  1.00027e+05  &  0.89158   &  -0.12236  &  1.01394   \\
(2)1/2(7)   &         $\leftarrow$  &         (1)7/2(11)  &         8.587    &       -2755.724   &       -641.854    &     -191.649    &  3.54980e+03  &  1.85585e+02  &  -0.65858  &  -0.16523  &  -0.49335  \\
(2)3/2(5)   &         $\leftarrow$  &         (2)1/2(4)   &         7.021    &       2113.148    &       601.917     &     59.189      &  9.98287e+04  &  3.50224e+03  &  0.25129   &  0.12670   &  0.12459   \\
(2)1/2(29)  &         $\leftarrow$  &         (1)5/2(32)  &         11.048   &       -3313.907   &       -599.891    &     -107.984    &  3.84803e+03  &  1.34853e+05  &  -0.55636  &  -0.19870  &  -0.35767  \\
(1)5/2(23)  &         $\leftarrow$  &         (2)1/2(20)  &         12.087   &       3482.635    &       576.250     &     99.660      &  1.30728e+05  &  3.73489e+03  &  0.56995   &  0.20881   &  0.36113   \\
(1)3/2(34)  &         $\leftarrow$  &         (1)7/2(28)  &         9.506    &       -2706.882   &       -569.535    &     225.752     &  3.02128e+02  &  3.38206e+02  &  0.48102   &  -0.16230  &  0.64333   \\
(3)3/2(3)   &         $\leftarrow$  &         (X)5/2(42)  &         160.313  &       -42551.753  &       -530.858    &     -92.171     &  1.25210e+07  &  3.67726e+02  &  -6.98116  &  -2.55134  &  -4.42982  \\
(X)5/2(45)  &         $\leftarrow$  &         (3)3/2(5)   &         160.598  &       42500.243   &       529.275     &     93.007      &  3.83155e+02  &  1.26751e+07  &  7.02617   &  2.54825   &  4.47792   \\
(3)3/2(4)   &         $\leftarrow$  &         (2)1/2(31)  &         165.790  &       -43291.036  &       -522.239    &     -62.499     &  1.25975e+07  &  3.87156e+03  &  -5.70201  &  -2.59567  &  -3.10634  \\
(2)1/2(12)  &         $\leftarrow$  &         (2)3/2(13)  &         7.864    &       -2036.506   &       -517.964    &     -51.904     &  3.62470e+03  &  9.99854e+04  &  -0.24447  &  -0.12211  &  -0.12236  \\
(3)3/2(4)   &         $\leftarrow$  &         (1)5/2(34)  &         182.423  &       -46567.874  &       -510.548    &     -63.325     &  1.25975e+07  &  1.35753e+05  &  -6.25531  &  -2.79214  &  -3.46317  \\
(2)3/2(8)   &         $\leftarrow$  &         (X)5/2(18)  &         10.401   &       2652.633    &       510.079     &     -384.065    &  9.98994e+04  &  1.91805e+02  &  -1.03851  &  0.15905   &  -1.19755  \\
(1)5/2(30)  &         $\leftarrow$  &         (X)5/2(39)  &         16.588   &       4044.458    &       487.634     &     -193.212    &  1.33910e+05  &  3.47116e+02  &  -0.71834  &  0.24250   &  -0.96084  \\
(2)1/2(30)  &         $\leftarrow$  &         (1)5/2(33)  &         13.546   &       -3295.091   &       -486.490    &     -87.979     &  3.85983e+03  &  1.35316e+05  &  -0.55486  &  -0.19757  &  -0.35729  \\
(2)3/2(4)   &         $\leftarrow$  &         (2)1/2(3)   &         8.865    &       2121.789    &       478.702     &     46.985      &  9.98013e+04  &  3.48604e+03  &  0.25209   &  0.12722   &  0.12487   \\
(1)5/2(22)  &         $\leftarrow$  &         (2)1/2(19)  &         15.997   &       3502.506    &       437.888     &     75.424      &  1.30311e+05  &  3.72163e+03  &  0.57173   &  0.21000   &  0.36172   \\
(X)5/2(40)  &         $\leftarrow$  &         (3)3/2(1)   &         198.068  &       42594.620   &       430.102     &     75.507      &  3.54457e+02  &  1.23672e+07  &  7.03745   &  2.55391   &  4.48354   \\
(3)3/2(4)   &         $\leftarrow$  &         (2)3/2(32)  &         210.017  &       -45130.499  &       -429.779    &     -51.194     &  1.25975e+07  &  9.97755e+04  &  -5.92921  &  -2.70596  &  -3.22325  \\
(X)5/2(39)  &         $\leftarrow$  &         (2)3/2(28)  &         11.984   &       -2573.244   &       -429.442    &     334.259     &  3.47116e+02  &  9.99838e+04  &  1.04662   &  -0.15429  &  1.20091   \\
(X)5/2(14)  &         $\leftarrow$  &         (1)5/2(6)   &         20.356   &       -4328.081   &       -425.241    &     158.357     &  1.54073e+02  &  1.26292e+05  &  0.70688   &  -0.25951  &  0.96638   \\
(2)1/2(13)  &         $\leftarrow$  &         (2)3/2(14)  &         9.631    &       -2026.108   &       -420.736    &     -42.286     &  3.63904e+03  &  9.99965e+04  &  -0.24358  &  -0.12148  &  -0.12209  \\
(X)5/2(38)  &         $\leftarrow$  &         (1)5/2(29)  &         19.301   &       -4053.932   &       -420.064    &     166.980     &  3.39839e+02  &  1.33440e+05  &  0.72315   &  -0.24307  &  0.96622   \\
(2)3/2(0)   &         $\leftarrow$  &         (1)1/2(6)   &         23.082   &       4817.993    &       417.464     &     -107.296    &  9.96746e+04  &  2.56563e+02  &  -0.45359  &  0.28888   &  -0.74247  \\
\bottomrule
\end{tabular}
\end{threeparttable}
}
\end{table*}

\end{document}

\title{Supplementary Material for: Enhanced sensitivity to variations of fundamental constants in highly charged molecules from 
analytic perturbation theory}
\date{\today}
\author{Carsten Z\"ulch}
\affiliation{Fachbereich Chemie, Philipps-Universit\"{a}t Marburg, Hans-Meerwein-Stra\ss{}e 4, 35032 Marburg}
\author{Konstantin Gaul}
\affiliation{Fachbereich Chemie, Philipps-Universit\"{a}t Marburg, Hans-Meerwein-Stra\ss{}e 4, 35032 Marburg}
\author{Robert Berger}
\email[]{robert.berger@uni-marburg.de}
\affiliation{Fachbereich Chemie, Philipps-Universit\"{a}t Marburg, Hans-Meerwein-Stra\ss{}e 4, 35032 Marburg}

\maketitle

\section{Description}%
Herein, we display the potential energy curve of \ce{CeF^2+} at the level of 2c-ZORA-cGHF, and
report numerical data used in various figures and tables of the
main text. Except for table S4, we give raw data obtained from the outputs of the calculations, which
provide more digits than we consider numerically stable (see computational details of
the main text).

\section{Potential energy curves}

\begin{figure}[!htb]
  \includegraphics[width=0.45\textwidth]{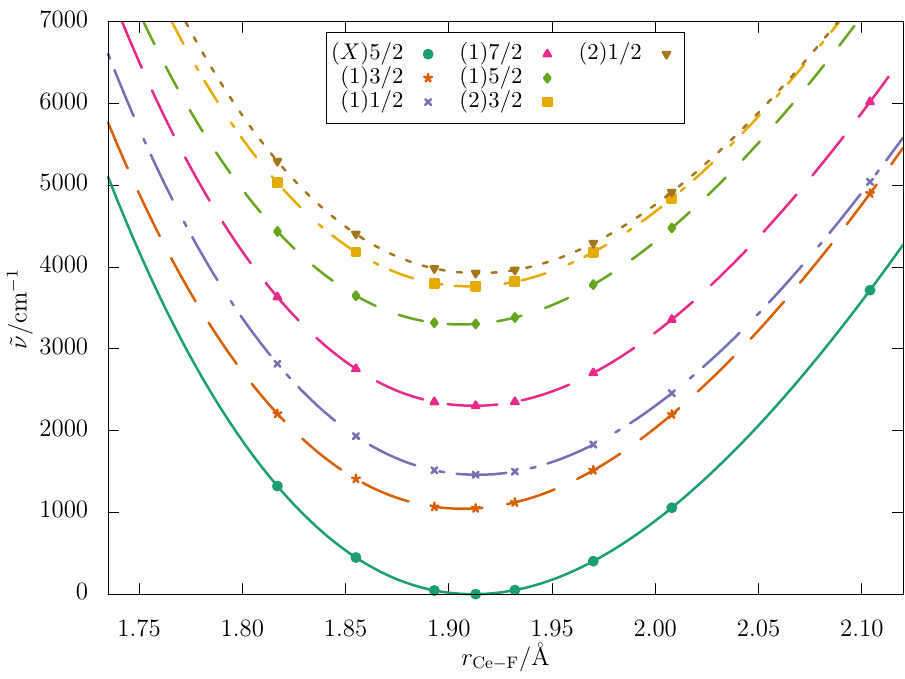}
  \caption{Potential energy curves (at
  $c=c_0=137.0359895~a_0E_\mathrm{h}\hbar^{-1}$) for the seven energetically
  lowest electronic states relative to the energetic minimum of the electronic
  ground state of \ce{CeF^2+} computed on the level of ZORA-cGHF.}
   \label{fig:pot1}
\end{figure}

\section{Tables of numerical data}

\begin{table}[!htb]
\begin{threeparttable}
\footnotesize
\caption{Internuclear distance of \ce{CeF^2+} used as equilibrium bond lengths for each 
electronic state (at $c=c_0=137.0359895~a_0E_\mathrm{h}\hbar^{-1}$) as described 
in the computational details.
}
\label{tab:re}
\begin{tabular}{
l
S[table-format=-2.15]
S[table-format=-2.11]
}
\toprule
{State}&
{$r_\mathrm{e}^\mathrm{ZORA-cGHF}\, /\, a_0$} &
{$r_\mathrm{e}^\mathrm{AOC-DHF}\, /\, a_0$} \\
\midrule
{$(\mathrm{X})5/2$} & 3.61312547638788 & 3.6135298110\\
{$(1)3/2$}          & 3.60338638092978 & 3.6032948440\\
{$(1)1/2$}          & 3.61684895933093 & 3.6167868380\\
{$(1)7/2$}          & 3.61346112265898 & 3.6136516840\\
{$(1)5/2$}          & 3.60142412482650 & 3.6018403390\\
{$(2)3/2$}          & 3.61033236642355 & 3.6106053300\\
{$(2)1/2$}          & 3.61761166283608 & 3.6187283850\\
\bottomrule
\end{tabular}
\end{threeparttable}
\end{table}

\begin{table}[!htb]
\resizebox{\textwidth}{!}{%
\begin{threeparttable}
\footnotesize
\caption{
Total energies of \ce{CeF^2+} (at $c=c_0=137.0359895~a_0E_\mathrm{h}\hbar^{-1}$)
as a function of the Ce--F internuclear distance calculated on the level of
2c-ZORA-cGHF. Values with '\textemdash' only converged to a less than desired
criteria and, thus, were not taken into account. 
}
\begin{tabular}{
S[table-format=1.5 ]
S[table-format=-4.11]
S[table-format=-4.11]
S[table-format=-4.11]
S[table-format=-4.11]
S[table-format=-4.11]
S[table-format=-4.11]
S[table-format=-4.11]
}
\toprule
{$r / a_0$}&
\multicolumn{7}{c}{$E/E_\text{h}$      }\\
&
{$(\mathrm{X})5/2$      }&
{$(1)3/2$      }&
{$(1)1/2$      }&
{$(1)7/2$      }&
{$(1)5/2$      }&
{$(2)3/2$      }&
{$(2)1/2$      }\\
\midrule
2.53005 & -9122.7114384697 & -9122.7170141680 & -9122.6998934368 & -9122.6935148032 & -9122.7008063410 & -9122.6920459975 & -9122.6891005953\\ 
2.71077 & -9122.8922703330 & -9122.8965940749 & -9122.8845134510 & -9122.8792697472 & -9122.8845129908 & -9122.8764437193 & -9122.8732492547\\
2.89149 & -9123.0095951165 & -9123.0112013262 & -9123.0028013120 & -9122.9982742922 & -9123.0006807690 & -9122.9942866340 & -9122.9913660838\\
3.07221 & -9123.0837556021 & -9123.0828005539 & -9123.0769725015 & -9123.0729382210 & -9123.0726965651 & -9123.0679158726 & -9123.0655486089\\
3.25292 & -9123.1272734137 & -9123.1244629227 & -9123.1204198579 & -9123.1166529071 & -9123.1143888139 & -9123.1108224260 & -9123.1090657608\\
3.43364 & -9123.1488620145 & -9123.1448284010 & -9123.1420441875 & -9123.1383267992 & -9123.1346663401 & -9123.1319596064 & -9123.1307566540\\
3.50593 & -9123.1528427787 & -9123.1484613957 & -9123.1460759983 & -9123.1423258572 & -9123.1382535408 & -9123.1358160471 & -9123.1348096879\\
3.57822 & -9123.1546783745 & -9123.1500151986 & -9123.1479824982 & -9123.1441739910 & -9123.1397595652 & -9123.1375584935 & -9123.1367340861\\
3.61436 & -9123.1548840332 & -9123.1501019040 & -9123.1482305652 & -9123.1443842047 & -9123.1398220893 & -9123.1377286514 & -9123.1369899214\\
3.65050 & -9123.1546600100 & -9123.1497721710 & -9123.1480532848 & -9123.1441637806 & -9123.1394681379 & -9123.1374761101 & -9123.1368196659\\
3.72279 & -9123.1530468074 & -9123.1479834547 & -9123.1465455378 & -9123.1425553713 & -9123.1376311984 & -9123.1358254735 & -9123.1353239203\\
3.79508 & -9123.1500692394 & -9123.1448723841 & -9123.1436875593 & -9123.1395800075 & -9123.1344725535 & -9123.1328341928 & -9123.1324754991\\
3.97580 & -9123.1379416688 & -9123.1325551470 & -9123.1319058815 & -9123.1274501898 & -9123.1220408677 & -9123.1207566872 & -9123.1207091846\\
4.15651 & -9123.1210980507 & -9123.1156675542 & -9123.1154484410 & -9123.1105977532 & -9123.1050465733 & -9123.1040491450 & -9123.1042583243\\
4.33723 & -9123.1014110283 & -9123.0960267867 & -9123.0961603636 & -9123.0908988055 & -9123.0853072890 & -9123.0845486587 & -9123.0849711388\\
4.51795 & -9123.0802021217 & -9123.0749168159 & -9123.0753436913 & -9123.0696768655 & -9123.0641067417 & -9123.0635510295 & -9123.0641516153\\
4.69867 & -9123.0583858007 & -9123.0532270609 & -9123.0539001637 & -9123.0478476301 & -9123.0423336482 & -9123.0419533985 & -9123.0427025675\\
4.87939 & -9123.0365829497 & -9123.0315612147 & -9123.0324420865 & -9123.0260329720 & -9123.0205902103 & -9123.0203658464 & -9123.0212365935\\
5.06010 & -9123.0152151536 & -9123.0103233840 & -9123.0113786278 & -9123.0046571263 & -9122.9992744539 & -9122.9992050224 & -9123.0001610647\\
5.24082 & -9122.9943377615 & -9122.9896469342 & -9122.9908729887 & -9122.9837527122 & -9122.9786382815 & -9122.9783248394 & -9122.9795658780\\
5.42154 & -9122.9745767858 & -9122.9699854104 & -9122.9713251542 & -9122.9639894708 & -9122.9588473516 & -9122.9588785837 & -9122.9600882713\\
5.60226 & -9122.9556639855 & -9122.9511846423 & -9122.9526293487 & -9122.9450706047 & -9122.9399858909 & -9122.9401196820 & -9122.9413752889\\
5.78298 & -9122.9377335814 & -9122.9333584512 & -9122.9348898350 & -9122.9271348710 & -9122.9221038881 & -9122.9223127215 & -9122.9236131562\\
5.96369 & -9122.9208125580 & -9122.9165337391 & -9122.9181327454 & -9122.9102098852 & -9122.9052209836 & -9122.9054907376 & -9122.9068260502\\
6.14441 & -9122.9049050086 & -9122.9007242383 & -9122.9023666466 & -9122.8942997561 & -9122.8893383401 & -9122.8896622554 & -9122.8910186719\\
6.32513 & -9122.8900025564 & -9122.8859902242 & -9122.8876044865 & -9122.8793953646 & -9122.8744460588 & -9122.8748380131 & -9122.8761905989\\
6.50585 & -9122.8760785563 & {\textemdash}    & {\textemdash}    & -9122.8654780969 & {\textemdash}    & -9122.8611828323 & -9122.8624461473\\
6.68657 & -9122.8631291586 & -9122.8589519181 & -9122.8608269466 & -9122.8525269541 & -9122.8475465957 & -9122.8478025761 & -9122.8492983767\\
6.86728 & -9122.8511160649 & -9122.8470716855 & -9122.8489191436 & -9122.8405156972 & -9122.8354947704 & -9122.8358146626 & -9122.8371984135\\
7.04800 & -9122.8400223320 & -9122.8361074800 & -9122.8379465296 & -9122.8294244297 & -9122.8243394336 & -9122.8246845678 & -9122.8260114202\\
7.22872 & -9122.8298502230 & -9122.8261148005 & -9122.8279431996 & -9122.8192548731 & -9122.8140690006 & -9122.8144430143 & -9122.8156764848\\ 
\bottomrule
\end{tabular}
\end{threeparttable}
}
\end{table}

\begin{table}
\resizebox{\textwidth}{!}{%
\begin{threeparttable}
\footnotesize
\caption{
Values of $q_1$ of individual electronic states in \ce{CeF^2+} as a function of
the Ce--F internuclear distance calculated on the level of 2c-ZORA-cGHF (at
$c=c_0=137.0359895~a_0E_\mathrm{h}\hbar^{-1}$). Values with '\textemdash' only
converged to a less than desired criteria and, thus, were not taken into
account.  
}
\begin{tabular}{
S[table-format=1.5 ]
S[table-format=-1.8e+1]
S[table-format=-1.8e+1]
S[table-format=-1.8e+1]
S[table-format=-1.8e+1]
S[table-format=-1.8e+1]
S[table-format=-1.8e+1]
S[table-format=-1.8e+1]
}
\toprule
{$r / a_0$}&
\multicolumn{7}{c}{$hc_0 q_1/E_\text{h}$      }\\
&
{$(\mathrm{X})5/2$      }&
{$(1)3/2$      }&
{$(1)1/2$      }&
{$(1)7/2$      }&
{$(1)5/2$      }&
{$(2)3/2$      }&
{$(2)1/2$      }\\
\midrule
2.53005 & -0.51142447E+03 & -0.51142849E+03 & -0.51144152E+03 & -0.51141871E+03 & -0.51142311E+03 & -0.51142949E+03 & -0.51143679E+03 \\ 
2.71077 & -0.51144183E+03 & -0.51144287E+03 & -0.51145188E+03 & -0.51143418E+03 & -0.51143587E+03 & -0.51144015E+03 & -0.51144698E+03 \\
2.89149 & -0.51145540E+03 & -0.51145460E+03 & -0.51146079E+03 & -0.51144630E+03 & -0.51144618E+03 & -0.51144931E+03 & -0.51145495E+03 \\
3.07221 & -0.51146400E+03 & -0.51146306E+03 & -0.51146735E+03 & -0.51145477E+03 & -0.51145403E+03 & -0.51145611E+03 & -0.51146038E+03 \\
3.25292 & -0.51146948E+03 & -0.51146869E+03 & -0.51147167E+03 & -0.51146026E+03 & -0.51145943E+03 & -0.51146070E+03 & -0.51146377E+03 \\
3.43364 & -0.51147287E+03 & -0.51147226E+03 & -0.51147434E+03 & -0.51146366E+03 & -0.51146291E+03 & -0.51146365E+03 & -0.51146579E+03 \\
3.50593 & -0.51147383E+03 & -0.51147328E+03 & -0.51147508E+03 & -0.51146461E+03 & -0.51146391E+03 & -0.51146450E+03 & -0.51146634E+03 \\
3.57822 & -0.51147462E+03 & -0.51147412E+03 & -0.51147569E+03 & -0.51146540E+03 & -0.51146475E+03 & -0.51146520E+03 & -0.51146679E+03 \\
3.61436 & -0.51147496E+03 & -0.51147449E+03 & -0.51147595E+03 & -0.51146574E+03 & -0.51146511E+03 & -0.51146551E+03 & -0.51146698E+03 \\
3.65050 & -0.51147527E+03 & -0.51147483E+03 & -0.51147618E+03 & -0.51146605E+03 & -0.51146545E+03 & -0.51146579E+03 & -0.51146715E+03 \\
3.72279 & -0.51147581E+03 & -0.51147542E+03 & -0.51147659E+03 & -0.51146658E+03 & -0.51146603E+03 & -0.51146628E+03 & -0.51146745E+03 \\
3.79508 & -0.51147627E+03 & -0.51147592E+03 & -0.51147693E+03 & -0.51146704E+03 & -0.51146652E+03 & -0.51146670E+03 & -0.51146771E+03 \\
3.97580 & -0.51147714E+03 & -0.51147687E+03 & -0.51147756E+03 & -0.51146788E+03 & -0.51146746E+03 & -0.51146749E+03 & -0.51146818E+03 \\
4.15651 & -0.51147774E+03 & -0.51147754E+03 & -0.51147799E+03 & -0.51146847E+03 & -0.51146812E+03 & -0.51146804E+03 & -0.51146852E+03 \\
4.33723 & -0.51147819E+03 & -0.51147802E+03 & -0.51147831E+03 & -0.51146890E+03 & -0.51146860E+03 & -0.51146846E+03 & -0.51146879E+03 \\
4.51795 & -0.51147853E+03 & -0.51147839E+03 & -0.51147856E+03 & -0.51146922E+03 & -0.51146896E+03 & -0.51146878E+03 & -0.51146900E+03 \\
4.69867 & -0.51147879E+03 & -0.51147868E+03 & -0.51147876E+03 & -0.51146946E+03 & -0.51146924E+03 & -0.51146903E+03 & -0.51146918E+03 \\
4.87939 & -0.51147898E+03 & -0.51147889E+03 & -0.51147891E+03 & -0.51146964E+03 & -0.51146946E+03 & -0.51146922E+03 & -0.51146932E+03 \\
5.06010 & -0.51147910E+03 & -0.51147905E+03 & -0.51147902E+03 & -0.51146974E+03 & -0.51146961E+03 & -0.51146933E+03 & -0.51146944E+03 \\
5.24082 & -0.51147901E+03 & -0.51147888E+03 & -0.51147890E+03 & -0.51146963E+03 & -0.51146972E+03 & -0.51146884E+03 & -0.51146889E+03 \\
5.42154 & -0.51147929E+03 & -0.51147922E+03 & -0.51147915E+03 & -0.51146991E+03 & -0.51146978E+03 & -0.51146951E+03 & -0.51146953E+03 \\
5.60226 & -0.51147929E+03 & -0.51147924E+03 & -0.51147916E+03 & -0.51146991E+03 & -0.51146979E+03 & -0.51146955E+03 & -0.51146955E+03 \\
5.78298 & -0.51147923E+03 & -0.51147920E+03 & -0.51147912E+03 & -0.51146984E+03 & -0.51146975E+03 & -0.51146951E+03 & -0.51146950E+03 \\
5.96369 & -0.51147911E+03 & -0.51147910E+03 & -0.51147903E+03 & -0.51146971E+03 & -0.51146964E+03 & -0.51146941E+03 & -0.51146940E+03 \\
6.14441 & -0.51147891E+03 & -0.51147890E+03 & -0.51147887E+03 & -0.51146951E+03 & -0.51146946E+03 & -0.51146923E+03 & -0.51146924E+03 \\
6.32513 & -0.51147863E+03 & -0.51147840E+03 & -0.51147859E+03 & -0.51146923E+03 & -0.51146922E+03 & -0.51146892E+03 & -0.51146898E+03 \\
6.50585 & -0.51147827E+03 & {\textemdash}   & {\textemdash}   & -0.51146888E+03 & {\textemdash}   & -0.51146795E+03 & -0.51146814E+03 \\
6.68657 & -0.51147787E+03 & -0.51147785E+03 & -0.51147799E+03 & -0.51146846E+03 & -0.51146853E+03 & -0.51146806E+03 & -0.51146773E+03 \\
6.86728 & -0.51147739E+03 & -0.51147754E+03 & -0.51147761E+03 & -0.51146798E+03 & -0.51146809E+03 & -0.51146790E+03 & -0.51146793E+03 \\
7.04800 & -0.51147684E+03 & -0.51147704E+03 & -0.51147712E+03 & -0.51146744E+03 & -0.51146759E+03 & -0.51146740E+03 & -0.51146743E+03 \\
7.22872 & -0.51147620E+03 & -0.51147640E+03 & -0.51147650E+03 & -0.51146680E+03 & -0.51146702E+03 & -0.51146679E+03 & -0.51146686E+03 \\
\bottomrule
\end{tabular}
\end{threeparttable}
}
\end{table}

\begin{table*}
\begin{threeparttable}
\caption{Electronic transitions in \ce{CeF^2+} and \ce{PaF^3+} sorted by their
absolute sensitivity $\Delta q_1$ to $\alpha$ variations as
computed on the level of ZORA-cGHF.}
\label{tab:vibstates_q}
\begin{tabular}{
l
c
l
S[table-format=6.0,round-mode=figures,round-precision=2]
S[table-format=-5.0,round-mode=figures,round-precision=2]
r
l
c
l
S[table-format=6.0,round-mode=figures,round-precision=2]
S[table-format=-5.0,round-mode=figures,round-precision=2]
}
\toprule
\multicolumn{5}{c}{\ce{CeF^2+}} & \phantom{aaaaaa}&
\multicolumn{5}{c}{\ce{PaF^3+}}
\\
\multicolumn{3}{c}{Transition} &
{$\Delta\tilde{\nu} / \si{cm^{-1}}$}  &
{$\Delta q_{1} / \si{cm^{-1}}$} &
&
\multicolumn{3}{c}{Transition} &
{$\Delta\tilde{\nu} /\si{cm^{-1}}$}  &
{$\Delta q_{1} / \si{cm^{-1}}$}  \\
\midrule
(3)3/2 & $\leftarrow$ & (1)5/2    &  28053.654    & -26051.128 &     &  (3)3/2 & $\leftarrow$ & (1)5/2   &  23306.822    & -46647.919      \\
(3)3/2 & $\leftarrow$ & (2)3/2    &  27589.468    & -25938.700 &     &  (3)3/2 & $\leftarrow$ & (1)7/2   &  24209.188    & -45570.119      \\
(3)3/2 & $\leftarrow$ & (1)7/2    &  29047.476    & -25887.555 &     &  (3)3/2 & $\leftarrow$ & (2)3/2   &  21735.730    & -44963.738      \\
(3)3/2 & $\leftarrow$ & (2)1/2    &  27428.316    & -25608.976 &     &  (3)3/2 & $\leftarrow$ & (1)3/2   &  28493.481    & -43013.050      \\
(3)3/2 & $\leftarrow$ & (1)3/2    &  30308.852    & -23988.047 &     &  (3)3/2 & $\leftarrow$ & (2)1/2   &  20912.376    & -42835.993      \\
(3)3/2 & $\leftarrow$ & (X)5/2    &  31352.218    & -23865.644 &     &  (3)3/2 & $\leftarrow$ & (X)5/2   &  29733.852    & -41931.745      \\
(3)3/2 & $\leftarrow$ & (1)1/2    &  29895.304    & -23642.309 &     &  (3)3/2 & $\leftarrow$ & (1)1/2   &  26713.553    & -40023.061      \\
(1)5/2 & $\leftarrow$ & (1)1/2    &   1841.650    &   2408.818 &     &  (1)5/2 & $\leftarrow$ & (1)1/2   &   3406.731    &   6624.858      \\
(2)3/2 & $\leftarrow$ & (1)1/2    &   2305.836    &   2296.390 &     &  (1)7/2 & $\leftarrow$ & (1)1/2   &   2504.366    &   5547.058      \\
(1)7/2 & $\leftarrow$ & (1)1/2    &    847.828    &   2245.246 &     &  (2)3/2 & $\leftarrow$ & (1)1/2   &   4977.823    &   4940.677      \\
(1)5/2 & $\leftarrow$ & (X)5/2    &   3298.564    &   2185.484 &     &  (1)5/2 & $\leftarrow$ & (X)5/2   &   6427.030    &   4716.174      \\
(2)3/2 & $\leftarrow$ & (X)5/2    &   3762.750    &   2073.056 &     &  (2)1/2 & $\leftarrow$ & (1)5/2   &   2394.446    &  -3811.926      \\
(1)5/2 & $\leftarrow$ & (1)3/2    &   2255.198    &   2063.080 &     &  (1)7/2 & $\leftarrow$ & (X)5/2   &   5524.664    &   3638.374      \\
(1)7/2 & $\leftarrow$ & (X)5/2    &   2304.742    &   2021.911 &     &  (1)5/2 & $\leftarrow$ & (1)3/2   &   5186.659    &   3634.869      \\
(2)1/2 & $\leftarrow$ & (1)1/2    &   2466.989    &   1966.667 &     &  (2)3/2 & $\leftarrow$ & (X)5/2   &   7998.121    &   3031.993      \\
(2)3/2 & $\leftarrow$ & (1)3/2    &   2719.384    &   1950.652 &     &  (1)1/2 & $\leftarrow$ & (1)3/2   &   1779.927    &  -2989.989      \\
(1)7/2 & $\leftarrow$ & (1)3/2    &   1261.376    &   1899.508 &     &  (2)1/2 & $\leftarrow$ & (1)1/2   &   5801.177    &   2812.931      \\
(2)1/2 & $\leftarrow$ & (X)5/2    &   3923.903    &   1743.332 &     &  (2)1/2 & $\leftarrow$ & (1)7/2   &   3296.812    &  -2734.126      \\
(2)1/2 & $\leftarrow$ & (1)3/2    &   2880.537    &   1620.929 &     &  (1)7/2 & $\leftarrow$ & (1)3/2   &   4284.293    &   2557.069      \\
(2)1/2 & $\leftarrow$ & (1)5/2    &    625.339    &   -442.152 &     &  (2)1/2 & $\leftarrow$ & (2)3/2   &    823.355    &  -2127.746      \\
(1)1/2 & $\leftarrow$ & (1)3/2    &    413.548    &   -345.738 &     &  (2)3/2 & $\leftarrow$ & (1)3/2   &   6757.750    &   1950.689      \\
(2)1/2 & $\leftarrow$ & (2)3/2    &    161.152    &   -329.724 &     &  (1)1/2 & $\leftarrow$ & (X)5/2   &   3020.298    &  -1908.684      \\
(2)1/2 & $\leftarrow$ & (1)7/2    &   1619.161    &   -278.579 &     &  (2)3/2 & $\leftarrow$ & (1)5/2   &   1571.091    &  -1684.181      \\
(1)1/2 & $\leftarrow$ & (X)5/2    &   1456.914    &   -223.334 &     &  (1)3/2 & $\leftarrow$ & (X)5/2   &   1240.371    &   1081.304      \\
(1)5/2 & $\leftarrow$ & (1)7/2    &    993.822    &    163.572 &     &  (1)5/2 & $\leftarrow$ & (1)7/2   &    902.366    &   1077.800      \\
(1)3/2 & $\leftarrow$ & (X)5/2    &   1043.366    &    122.404 &     &  (2)1/2 & $\leftarrow$ & (X)5/2   &   8821.476    &    904.247      \\
(2)3/2 & $\leftarrow$ & (1)5/2    &    464.186    &   -112.428 &     &  (2)3/2 & $\leftarrow$ & (1)7/2   &   2473.457    &   -606.381      \\
(2)3/2 & $\leftarrow$ & (1)7/2    &   1458.008    &     51.145 &     &  (2)1/2 & $\leftarrow$ & (1)3/2   &   7581.105    &   -177.057      \\
\bottomrule
\end{tabular}
\end{threeparttable}
\end{table*}

\begin{table}
\begin{threeparttable}
\footnotesize
\caption{ Values $q_1$, $2\left< H^{(2)} \right>$, $\left<\left< H^{(1)};H^{(1)}
\right>\right>$ for individual electronic states at the respective equilibrium bond
lengths of \ce{PaF^3+} in the energetically lowest eight electronic states 
(at $c=c_0=137.0359895~a_0E_\mathrm{h}\hbar^{-1}$). As
per equation (14) of the main text $q_2$ corresponds to the sum of an
expectation value $\left< H^{(2)} \right>$ and a response
contribution $\frac{1}{2}\left<\left< H^{(1)};H^{(1)} \right>\right>$, here, with the
Hamiltonians given in equation (24) and (25).
}
\begin{tabular}{
l
S[table-format=-1.15E1]
S[table-format=-1.14E1]
S[table-format=-1.15E1]
S[table-format=-1.15E1]
S[table-format=-1.15E1]
S[table-format=-1.15E1]
S[table-format=-1.15E1]
}
\toprule
{State}&
{$hc_0 q_1/E_\mathrm{h}$} &
{$2\left< H^{(2)} \right>/E_\mathrm{h}$} &
{$\left<\left< H^{(1)};H^{(1)} \right>\right>/E_\mathrm{h}$} \\
\midrule
{$(\mathrm{X})5/2$} & -0.496949169591207E+04 & 0.30318055906334e+04 & -0.693321864122968E+04 \\ 
{$(1)3/2$} & -0.496948833200438E+04 & 0.30318033376218e+04 & -0.693322288613604E+04 \\
{$(1)1/2$} & -0.496950127145434E+04 & 0.30318074424279e+04 & -0.693324494079290E+04 \\
{$(1)7/2$} & -0.496947586927064E+04 & 0.30318102902769e+04 & -0.693324571117269E+04 \\
{$(1)5/2$} & -0.496947191574450E+04 & 0.30318080009806e+04 & -0.693324677363224E+04 \\
{$(2)3/2$} & -0.496947890869696E+04 & 0.30318099541235e+04 & -0.693326520274510E+04 \\
{$(2)1/2$} & -0.496948855155430E+04 & 0.30318140311151e+04 & -0.693327072658838E+04 \\
{$(3)3/2$} & -0.496968688624922E+04 & 0.30318807887135e+04 & -0.693330099683591E+04 \\
\bottomrule
\end{tabular}
\end{threeparttable}
\end{table}

\begin{table}
\begin{threeparttable}
\footnotesize
\caption{ Values $q_1$, $2\left< H^{(2)} \right>$, $\left<\left< H^{(1)};H^{(1)}
\right>\right>$ for individual electronic states at the respective equilibrium bond
lengths of \ce{CeF^2+} in the energetically lowest eight electronic states
(at $c=c_0=137.0359895~a_0E_\mathrm{h}\hbar^{-1}$). As
per equation (14) of the main text $q_2$ corresponds to the sum of an
expectation value $\left< H^{(2)} \right>$ and a response
contribution $\frac{1}{2}\left<\left< H^{(1)};H^{(1)} \right>\right>$, here, with the
Hamiltonians given in equation (24) and (25).
}
\begin{tabular}{
l
S[table-format=-3.13]
S[table-format=-3.13]
S[table-format=-3.13]
S[table-format=-3.13]
S[table-format=-3.13]
S[table-format=-3.13]
S[table-format=-3.13]
}
\toprule
{State}&
{$hc_0 q_1/E_\mathrm{h}$} &
{$2\left< H^{(2)} \right>/E_\mathrm{h}$} &
{$\left<\left< H^{(1)};H^{(1)} \right>\right>/E_\mathrm{h}$} \\
\midrule
{$(\mathrm{X})5/2$} & -511.474947005129 & 67.3120105007154 & -258.195512678799 \\
{$(1)3/2$} & -511.474385189328 & 67.3119078582840 & -258.195451471348 \\
{$(1)1/2$} & -511.475962842081 & 67.3120058911547 & -258.196744264992 \\
{$(1)7/2$} & -511.465728411338 & 67.3121691650772 & -258.199141085717 \\
{$(1)5/2$} & -511.464985708549 & 67.3120686589679 & -258.198971946710 \\
{$(2)3/2$} & -511.465477591613 & 67.3120921877509 & -258.199627535626 \\
{$(2)1/2$} & -511.466997502706 & 67.3122226891123 & -258.200485072514 \\
\bottomrule
\end{tabular}
\end{threeparttable}
\end{table}

\begin{table}
\resizebox{\textwidth}{!}{%
\begin{threeparttable}
\footnotesize
\caption{Total energies of \ce{CeF^2+} as a function of the speed of light $c$,
with $c_0=137.0359895~a_0\,E_\text{h}\,\hbar^{-1}$, calculated on the level of
2c-ZORA-cGHF with the internuclear distances for each electronic state reported 
in Tab.~\ref{tab:re}.}
\begin{tabular}{
S[table-format=3.7 ]
S[table-format=-4.11]
S[table-format=-4.11]
S[table-format=-4.11]
S[table-format=-4.11]
S[table-format=-4.11]
S[table-format=-4.11]
S[table-format=-4.11]
}
\toprule
{$c/(a_0 E_\text{h} \hbar^{-1})$}&
\multicolumn{7}{c}{$E/E_\text{h}$      }\\
&
{$(\mathrm{X})5/2$      }&
{$(1)3/2$      }&
{$(1)1/2$      }&
{$(1)7/2$      }&
{$(1)5/2$      }&
{$(2)3/2$      }&
{$(2)1/2$      }\\
\midrule
123.0359895 & -9249.8796018203 & -9249.8747104323 & -9249.8732382756 & -9249.8669829156 & -9249.8622717964 & -9249.8602966307 & -9249.8599432484 \\ 
124.0359895 & -9239.0926911253 & -9239.0878095993 & -9239.0862995420 & -9239.0802400875 & -9239.0755432132 & -9239.0735538269 & -9239.0731666413 \\
125.0359895 & -9228.6156955713 & -9228.6108238439 & -9228.6092774609 & -9228.6034097440 & -9228.5987268584 & -9228.5967240936 & -9228.5963042707 \\
126.0359895 & -9218.4363498632 & -9218.4314878513 & -9218.4299065942 & -9218.4242266187 & -9218.4195574583 & -9218.4175420908 & -9218.4170907278 \\
127.0359895 & -9208.5430124448 & -9208.5381600531 & -9208.5365453114 & -9208.5310491774 & -9208.5263934815 & -9208.5243662193 & -9208.5238843585 \\
128.0359895 & -9198.9246265807 & -9198.9197837017 & -9198.9181367691 & -9198.9128207138 & -9198.9081782205 & -9198.9061397104 & -9198.9056283420 \\
129.0359895 & -9189.5706843195 & -9189.5658508356 & -9189.5641729280 & -9189.5590333071 & -9189.5544037547 & -9189.5523545878 & -9189.5518146540 \\
130.0359895 & -9180.4711930926 & -9180.4663688790 & -9180.4646611415 & -9180.4596944176 & -9180.4550775477 & -9180.4530182803 & -9180.4524506630 \\
131.0359895 & -9171.6166447233 & -9171.6118296475 & -9171.6100931601 & -9171.6052959037 & -9171.6006914574 & -9171.5986225775 & -9171.5980281404 \\
132.0359895 & -9162.9979866475 & -9162.9931805733 & -9162.9914163539 & -9162.9867852304 & -9162.9821929560 & -9162.9801149271 & -9162.9794944766 \\
133.0359895 & -9154.6065951622 & -9154.6017979474 & -9154.6000069633 & -9154.5955387343 & -9154.5909583771 & -9154.5888716261 & -9154.5882259313 \\
134.0359895 & -9146.4342505359 & -9146.4294620383 & -9146.4276451932 & -9146.4233367111 & -9146.4187680278 & -9146.4166729451 & -9146.4160027429 \\
135.0359895 & -9138.4731138327 & -9138.4683339038 & -9138.4664920641 & -9138.4623402637 & -9138.4577830110 & -9138.4556799563 & -9138.4549859510 \\
136.0359895 & -9130.7157053120 & -9130.7109338065 & -9130.7090677897 & -9130.7050696895 & -9130.7005236257 & -9130.6984129446 & -9130.6976957956 \\
136.6359895 & -9126.1560656669 & -9126.1512991438 & -9126.1494189934 & -9126.1455117144 & -9126.1409722538 & -9126.1388571334 & -9126.1381264212 \\
136.7359895 & -9125.4029007251 & -9125.3981350218 & -9125.3962525485 & -9125.3923603074 & -9125.3878219372 & -9125.3857060859 & -9125.3849731366 \\
136.8359895 & -9124.6516533359 & -9124.6468884622 & -9124.6450036545 & -9124.6411264280 & -9124.6365891480 & -9124.6344725799 & -9124.6337373886 \\
136.9359895 & -9123.9023167579 & -9123.8975527061 & -9123.8956655902 & -9123.8918033374 & -9123.8872671454 & -9123.8851498473 & -9123.8844124364 \\
137.0359895 & -9123.1548842807 & -9123.1501210497 & -9123.1482316313 & -9123.1443843260 & -9123.1398492190 & -9123.1377312075 & -9123.1369915692 \\
137.1359895 & -9122.4093492245 & -9122.4045868137 & -9122.4026950861 & -9122.3988627123 & -9122.3943286886 & -9122.3922099647 & -9122.3914681070 \\
137.2359895 & -9121.6657049390 & -9121.6609433445 & -9121.6590493219 & -9121.6552318490 & -9121.6506989058 & -9121.6485794612 & -9121.6478354001 \\
137.3359895 & -9120.9239448049 & -9120.9191840271 & -9120.9172877236 & -9120.9134851126 & -9120.9089532468 & -9120.9068330915 & -9120.9060868288 \\
137.4359895 & -9120.1840622323 & -9120.1793022684 & -9120.1774036979 & -9120.1736159157 & -9120.1690851262 & -9120.1669642609 & -9120.1662158032 \\
138.0359895 & -9115.7838302852 & -9115.7790751781 & -9115.7771630969 & -9115.7734637602 & -9115.7689393773 & -9115.7668143306 & -9115.7660528076 \\
139.0359895 & -9108.5960255393 & -9108.5912784088 & -9108.5893443645 & -9108.5857902384 & -9108.5812763549 & -9108.5791445282 & -9108.5783617180 \\
140.0359895 & -9101.5852384941 & -9101.5804991929 & -9101.5785438601 & -9101.5751322500 & -9101.5706286440 & -9101.5684902917 & -9101.5676867485 \\
141.0359895 & -9094.7455084661 & -9094.7407768442 & -9094.7388008531 & -9094.7355291474 & -9094.7310356014 & -9094.7288909521 & -9094.7280672151 \\
142.0359895 & -9088.0711312417 & -9088.0664071571 & -9088.0644111153 & -9088.0612767499 & -9088.0567930538 & -9088.0546423234 & -9088.0537989066 \\
143.0359895 & -9081.5566455878 & -9081.5519288937 & -9081.5499133838 & -9081.5469138658 & -9081.5424398084 & -9081.5402831973 & -9081.5394205943 \\
144.0359895 & -9075.1968206034 & -9075.1921111559 & -9075.1900767301 & -9075.1872096261 & -9075.1827450048 & -9075.1805826952 & -9075.1797013841 \\
145.0359895 & -9068.9866438506 & -9068.9819415107 & -9068.9798887002 & -9068.9771516291 & -9068.9726962444 & -9068.9705284133 & -9068.9696288471 \\
146.0359895 & -9062.9213102089 & -9062.9166148359 & -9062.9145441489 & -9062.9119347861 & -9062.9074884460 & -9062.9053152534 & -9062.9043978717 \\
147.0359895 & -9056.9962114033 & -9056.9915228622 & -9056.9894347802 & -9056.9869508567 & -9056.9825133740 & -9056.9803349723 & -9056.9794001961 \\
148.0359895 & -9051.2069261582 & -9051.2022443134 & -9051.2001393001 & -9051.1977786054 & -9051.1933497917 & -9051.1911663237 & -9051.1902145553 \\
149.0359895 & -9045.5492109339 & -9045.5445356523 & -9045.5424141585 & -9045.5401745255 & -9045.5357541960 & -9045.5335657950 & -9045.5325974259 \\
150.0359895 & -9040.0189912063 & -9040.0143223583 & -9040.0121848081 & -9040.0100641228 & -9040.0056521023 & -9040.0034588932 & -9040.0024742973 \\
151.0359895 & -9034.6123532528 & -9034.6076907106 & -9034.6055375174 & -9034.6035337059 & -9034.5991298256 & -9034.5969319239 & -9034.5959314628 \\
\bottomrule
\end{tabular}
\end{threeparttable}
}
\end{table}

\begin{table}
\resizebox{\textwidth}{!}{%
\begin{threeparttable}
\footnotesize
\caption{Total energies of \ce{CeF^2+} as a function of the speed of light $c$,
with $c_0=137.0359895~a_0\,E_\text{h}\,\hbar^{-1}$, calculated on the level of
AOC-DHF with the internuclear distances for each electronic state reported in
Tab.~\ref{tab:re}.}
\begin{tabular}{
S[table-format=3.7 ]
S[table-format=-4.13]
S[table-format=-4.13]
S[table-format=-4.13]
S[table-format=-4.13]
S[table-format=-4.13]
S[table-format=-4.13]
S[table-format=-4.13]
}
\toprule
{$c/(a_0 E_\text{h} \hbar^{-1})$}&
\multicolumn{7}{c}{$E/E_\text{h}$      }\\
&
{$(\mathrm{X})5/2$      }&
{$(1)3/2$      }&
{$(1)1/2$      }&
{$(1)7/2$      }&
{$(1)5/2$      }&
{$(2)3/2$      }&
{$(2)1/2$      }\\
\midrule
123.0359895 & -9040.3423033583495 & -9040.3366225461323 & -9040.3341268446475 & -9040.3306652848278 & -9040.3254092987281 & -9040.3228841108121 & -9040.3214629469894 \\
124.0359895 & -9033.5830748130338 & -9033.5774044013360 & -9033.5748699995838 & -9033.5715914878056 & -9033.5663493207867 & -9033.5638108113526 & -9033.5623517103813 \\
125.0359895 & -9027.0125277050847 & -9027.0068676261199 & -9027.0042962346834 & -9027.0011966890543 & -9026.9959680894026 & -9026.9934170737961 & -9026.9919213075082 \\
126.0359895 & -9020.6234333552347 & -9020.6177835261169 & -9020.6151767506744 & -9020.6122522290370 & -9020.6070369436075 & -9020.6044741673231 & -9020.6029429402297 \\
127.0359895 & -9014.4089183605010 & -9014.4032786878506 & -9014.4006380376177 & -9014.3978847288763 & -9014.3926825039380 & -9014.3901086492388 & -9014.3885431036670 \\
128.0359895 & -9008.3624431708940 & -9008.3568135509468 & -9008.3541404474508 & -9008.3515546627059 & -9008.3463652453356 & -9008.3437809382885 & -9008.3421821586016 \\
129.0359895 & -9002.4777821983917 & -9002.4721625192742 & -9002.4694583030196 & -9002.4670364692047 & -9002.4618596076707 & -9002.4592654222379 & -9002.4576344392408 \\
130.0359895 & -8996.7490053335787 & -8996.7433954771786 & -8996.7406614137599 & -8996.7384000664460 & -8996.7332355109411 & -8996.7306319764921 & -8996.7289697699343 \\
131.0359895 & -8991.1704607554457 & -8991.1648605990376 & -8991.1620978853989 & -8991.1599936626790 & -8991.1548411657968 & -8991.1522287684656 & -8991.1505362713378 \\
132.0359895 & -8985.7367589286659 & -8985.7311683451826 & -8985.7283781148490 & -8985.7264277524318 & -8985.7212870703697 & -8985.7186662588410 & -8985.7169443600269 \\
133.0359895 & -8980.4427576938215 & -8980.4371765530923 & -8980.4343598802898 & -8980.4325602064982 & -8980.4274310977398 & -8980.4248022865613 & -8980.4230518336626 \\
134.0359895 & -8975.2835483612398 & -8975.2779765315936 & -8975.2751344357566 & -8975.2734823671854 & -8975.2683645951747 & -8975.2657281669417 & -8975.2639499692559 \\
135.0359895 & -8970.2544427362154 & -8970.2488800838528 & -8970.2460135348392 & -8970.2445060706323 & -8970.2393994020404 & -8970.2367557117705 & -8970.2349505402381 \\
136.0359895 & -8965.3509609954508 & -8965.3454073866887 & -8965.3425173064206 & -8965.3411515269909 & -8965.3360557331325 & -8965.3334051094844 & -8965.3315737032117 \\
136.6359895 & -8962.4673819041145 & -8962.4618336572184 & -8962.4589298824940 & -8962.4576477454320 & -8962.4525583668383 & -8962.4499037307269 & -8962.4480569243551 \\
136.7359895 & -8961.9909689649303 & -8961.9854216072199 & -8961.9825155804047 & -8961.9812472845369 & -8961.9761589677291 & -8961.9735036736110 & -8961.9716543245340 \\
136.8359895 & -8961.5157404246111 & -8961.5101939547603 & -8961.5072856834649 & -8961.5060312022670 & -8961.5009439445548 & -8961.4982879947547 & -8961.4964361101775 \\
136.9359895 & -8961.0416922335407 & -8961.0361466491322 & -8961.0332361428063 & -8961.0319954480019 & -8961.0269092471281 & -8961.0242526452148 & -8961.0223982315383 \\
137.0359895 & -8960.5688203584723 & -8960.5632756594896 & -8960.5603629263369 & -8960.5591359892442 & -8960.5540508438844 & -8960.5513935916424 & -8960.5495366564046 \\
137.1359895 & -8960.0971207863076 & -8960.0915769704534 & -8960.0886620188321 & -8960.0874488120953 & -8960.0823647196376 & -8960.0797068202901 & -8960.0778473700739 \\
137.2359895 & -8959.6265895186571 & -8959.6210465856020 & -8959.6181294232047 & -8959.6169299196281 & -8959.6118468773257 & -8959.6091883341960 & -8959.6073263748203 \\
137.3359895 & -8959.1572225769687 & -8959.1516805243391 & -8959.1487611604407 & -8959.1475753327868 & -8959.1424933386734 & -8959.1398341542936 & -8959.1379696932891 \\
137.4359895 & -8958.6890159994509 & -8958.6834748262318 & -8958.6805532689068 & -8958.6793810890176 & -8958.6743001409086 & -8958.6716403179507 & -8958.6697733617038 \\
138.0359895 & -8955.9039244870528 & -8955.8983885633806 & -8955.8954540144350 & -8955.8943631493075 & -8955.8892884320012 & -8955.8866248349186 & -8955.8847430465667 \\
139.0359895 & -8951.3523535589375 & -8951.3468262759852 & -8951.3438707117148 & -8951.3429132198726 & -8951.3378487146729 & -8951.3351790361558 & -8951.3332730411657 \\
140.0359895 & -8946.9103549717147 & -8946.9048361985697 & -8946.9018603816694 & -8946.9010336311640 & -8946.8959791259313 & -8946.8933036114595 & -8946.8913740312437 \\
141.0359895 & -8942.5743346269883 & -8942.5688242309352 & -8942.5658288925715 & -8942.5651303178565 & -8942.5600856041492 & -8942.5574044834775 & -8942.5554519119542 \\
142.0359895 & -8938.3408487474971 & -8938.3353465995224 & -8938.3323324386565 & -8938.3317595345216 & -8938.3267244109520 & -8938.3240378965238 & -8938.3220629057942 \\
143.0359895 & -8934.2065962012221 & -8934.2011021730304 & -8934.1980698600546 & -8934.1976201836023 & -8934.1925944505529 & -8934.1899027420077 & -8934.1879058802751 \\
144.0359895 & -8930.1684112859621 & -8930.1629252513540 & -8930.1598754278548 & -8930.1595465937644 & -8930.1545300576017 & -8930.1518333422991 & -8930.1498151364904 \\
145.0359895 & -8926.2232569466960 & -8926.2177787810178 & -8926.2147120643494 & -8926.2145017441053 & -8926.2094942146850 & -8926.2067926673717 & -8926.2047536240425 \\
146.0359895 & -8922.3682183969831 & -8922.3627479782135 & -8922.3596649595329 & -8922.3595708793073 & -8922.3545721709106 & -8922.3518659555793 & -8922.3498065619642 \\
147.0359895 & -8918.6004971107886 & -8918.5950343185468 & -8918.5919355674687 & -8918.5919555052387 & -8918.5869654369180 & -8918.5842547072662 & -8918.5821754326571 \\
148.0359895 & -8914.9174051681021 & -8914.9119498845412 & -8914.9088359481702 & -8914.9089677337088 & -8914.9039861276797 & -8914.9012710288862 & -8914.8991723252111 \\
149.0359895 & -8911.3163599234595 & -8911.3109120326681 & -8911.3077834379237 & -8911.3080249504856 & -8911.3030516337731 & -8911.3003323014291 & -8911.2982146039849 \\
150.0359895 & -8907.7948789786678 & -8907.7894383674757 & -8907.7862956216395 & -8907.7866447880060 & -8907.7816795915587 & -8907.7789561539375 & -8907.7768198833601 \\
151.0359895 & -8904.3505754407615 & -8904.3451419978028 & -8904.3419855900811 & -8904.3424403836161 & -8904.3374831426445 & -8904.3347557201250 & -8904.3326012811303 \\
\bottomrule
\end{tabular}
\end{threeparttable}
}
\end{table}

\begin{table}
\resizebox{\textwidth}{!}{%
\begin{threeparttable}
\footnotesize
\caption{Total energies of \ce{CeF^2+} as a function of the speed of light $c$, with $c_0=137.0359895~a_0\,E_\text{h}\,\hbar^{-1}$, calculated on the
level of AOC-DHF-FSCCSD with the internuclear distances for each electronic state set to $r=3.5392944488~a_0$.}
\begin{tabular}{
S[table-format=3.7 ]
S[table-format=-4.12]
S[table-format=-4.12]
S[table-format=-4.12]
S[table-format=-4.12]
S[table-format=-4.12]
S[table-format=-4.12]
S[table-format=-4.12]
}
\toprule
{$c/(a_0 E_\text{h} \hbar^{-1})$}&
\multicolumn{7}{c}{$E/E_\text{h}$      }\\
&
{$(\mathrm{X})5/2$      }&
{$(1)3/2$      }&
{$(1)1/2$      }&
{$(1)7/2$      }&
{$(1)5/2$      }&
{$(2)3/2$      }&
{$(2)1/2$      }\\
\midrule
125.0359895 &  -9028.151955150770 &  -9028.149984892745 &  -9028.146073649370 &  -9028.140190409373 &  -9028.139212289434 &  -9028.135307322404 &  -9028.133774166785 \\ 
127.0359895 &  -9015.547829250103 &  -9015.545880203093 &  -9015.541879081566 &  -9015.536369438474 &  -9015.535416943314 &  -9015.531437045414 &  -9015.529871043738 \\
129.0359895 &  -9003.616189976461 &  -9003.614262644945 &  -9003.610180400170 &  -9003.605025160936 &  -9003.604097373001 &  -9003.600048257205 &  -9003.598455499407 \\
131.0359895 &  -8992.308378280259 &  -8992.306472932687 &  -8992.302317323702 &  -8992.297498795022 &  -8992.296594762449 &  -8992.292481321383 &  -8992.290867475960 \\
133.0359895 &  -8981.580197307636 &  -8981.578314035409 &  -8981.574091954695 &  -8981.569593766424 &  -8981.568712527940 &  -8981.564538968089 &  -8981.562909307240 \\
135.0359895 &  -8971.391416327288 &  -8971.389555089378 &  -8971.385272672980 &  -8971.381079630648 &  -8971.380220232888 &  -8971.375990184144 &  -8971.374349607568 \\
137.0359895 &  -8961.705339402562 &  -8961.703500059388 &  -8961.699162781169 &  -8961.695260741553 &  -8961.694422250799 &  -8961.690138856444 &  -8961.688491913626 \\
140.0359895 &  -8948.046212908133 &  -8948.044405992096 &  -8948.039995432337 &  -8948.036505420734 &  -8948.035696601446 &  -8948.031338197454 &  -8948.029689520481 \\
143.0359895 &  -8935.341816891256 &  -8935.340041763646 &  -8935.335567269574 &  -8935.332462387509 &  -8935.331681314081 &  -8935.327252917126 &  -8935.325610975133 \\
146.0359895 &  -8923.502824696099 &  -8923.501080621339 &  -8923.496550149617 &  -8923.493805940854 &  -8923.493050811632 &  -8923.488556536349 &  -8923.486928856701 \\
149.0359895 &  -8912.450373763002 &  -8912.448659955322 &  -8912.444080305491 &  -8912.441674446627 &  -8912.440943586882 &  -8912.436386831283 &  -8912.434780079955 \\
152.0359895 &  -8902.114595898040 &  -8902.112911552884 &  -8902.108288559182 &  -8902.106200601176 &  -8902.105492460731 &  -8902.100876050934 &  -8902.099296105351 \\
\bottomrule
\end{tabular}
\end{threeparttable}
}
\end{table}